# Critical Review of Theoretical Models for Anomalous Effects (Cold Fusion) in Deuterated Metals

**V.A. Chechin[1], V.A.Tsarev[1], M. Rabinowitz[2], and Y.E. Kim[3]**

Inquiries to: Mario Rabinowitz; E-mail: Mario715@earthlink.net

**Abstract**

We briefly summarize the reported anomalous effects in deuterated metals at ambient temperature, commonly known as "Cold Fusion" (CF), with an emphasis on important experiments as well as the theoretical basis for the opposition to interpreting them as cold fusion. Then we critically examine more than 25 theoretical models for CF, including unusual nuclear and exotic chemical hypotheses. We conclude that they do not explain the data.

[1]Lebedev Physical Institute, Russian Academy of Sciences, Russian Federation

[2]Electric Power Research Institute, and Armor Research, 715 Lakemead Way, Redwood City, CA 94062-3922

[3]Department of Physics, Purdue University, West Lafayette, IN 47907

## 1. INTRODUCTION

Cold Fusion (CF) appears to have burst upon the scene in 1989 with the announcements of fusion at room temperature (Fleischmann and Pons, 1989; Jones et al, 1989) in palladium (Pd) electrolytic cells using heavy water (deuterium oxide). Actually its history spans 67 years with the report by Paneth and Peters (1926) of fusing hydrogen into He in a high temperature Pd capillary tube. Even though they retracted their claim (Paneth and Peters, 1927), Tandberg filed for patent in 1927 in Sweden for an electrolytic cell similar to present-day cells. Since the 1970's the Soviets have shown an interest in the possibility of fusion in the solid state (Mamyrin et al, 1978) and have actively investigated various fusion schemes including electrolytic cell fusion by Alikin (Rabinowitz, 1990a).High pressure piezonuclear fusion at room temperature was considered by Van Siclen and Jones (1986).

Muon catalyzed fusion (MCF) as an accepted but quite different form of CF was first suggested in England by Frank (1947) to explain experimental anomalies. Here the electron in a deuterium atom is replaced by a muon. Since the orbital radius is $\propto 1/(mass)$, the muon orbit is 207 times smaller than the electron orbit. As we shall see in Sec. 4.6, different sorts of tight orbits are invoked as one kind of explanation for CF. Can MCF from cosmic ray muons account for the solidly established very low neutron (n) level observations of CF (cf. Secs.4.2.1, 4.2.5)? Or are the n's an artifact indirectly related to natural alpha decay of nuclei?

By now many books and reviews have been written on the subject of CF by both its opponents and proponents. We feel that there is a serious need for a balanced account. It is our goal to present a perspective that is as balanced and objective as possible. Where feasible, we will point out shortcomings in the theory and experiments both by advocates and adversaries. We will only briefly cover and update the worldwide phenomenology, as previous CF review papers have covered the older experimental results.

Our review attempts to fill the need for an in-depth critical theoretical inquiry that is equitable to both sides of the issue. The problem of an adequate theoretical model of CF has turned out to be no simpler than the problem of its unambiguous experimental proof. Our paper deals with the following essential issues:

1) Is CF real or is it just due to artifacts?
2) What are the true properties and nature of CF?
3) If CF is real, what theoretical explanation(s) apply?
4) If CF is real, is ordinary physics sufficient to explain the mechanism or is extraordinary physics necessary?

Even if CF is real, it is premature to try to answer two more relevant questions which need to be addressed when more is known:
5) What new knowledge and benefits can we accrue from CF?
6) Can practical applications be derived from CF?

Although our paper raises the above questions and deals earnestly with them, the present state of knowledge does not yet permit definitive answers to even the first four. We will attempt to answer them within the limits of our ability combined with the availability of reliable experimental and theoretical knowledge. Of the multitude of hypotheses and models that have been suggested, we shall primarily focus on those that have a quantitative aspect to them in addition to their qualitative perspective. This will permit us to analyze the ramifications of their predictions, their internal consistency, and their consistency with established experiments and theory in other domains.

The urgent need for a "reasonable model" is dictated not only by an obvious wish to understand the nature of the phenomenon. Because of the still uncertain experimental situation, a good trustworthy working hypothesis would be invaluable if it could correlate a large number of observations, make predictions, and stimulate experiments. Theoretical demonstration of CF "permissibility in principle" would also provide an important psychological factor for putting the reported phenomenon into a framework which might lead to its general acceptance. Although to date both the experimental and theoretical domains have their difficulties, we are motivated to understand CF in an earnest quest for the truth wherever it may lead.

## 2. THEORETICAL OPPOSITION AND SUPPORT FOR COLD FUSION

It now appears likely that CF is a sporadic, non-equilibrium process. The initial expectation of a considerable number of theoretical publications was that CF is a continuous process associated with steady state conditions in a lattice. In this context, considerations were given to the difference in interatomic fields in a solid than in a plasma due to electron screening. The solid lattice environment permits the mutual approach of free deuterons (d) to much closer distances than they could otherwise be at ambient temperature. Although the average separation of d's is about 1.4 Å in heavily loaded Pd, the d's can be in equilibrium at a separation as close as 0.94 Å. (Sun and Tomanek, 1989; Liu et al, 1989; Lohr, 1989). Though this is closer than the 1.11Å separation of d's in $D_2^+$, this is not as close as the 0.74Å in $D_2$ shown schematically in Fig.1, which gives no measurable fusion rate. However, closer separation may be possible in non-equilibrium processes.

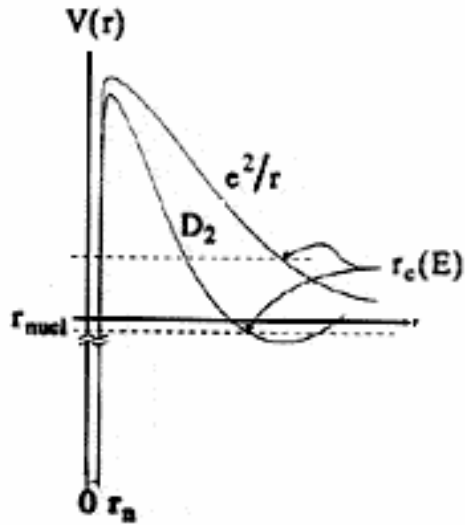

**Figure 1.** The Coulomb barrier and nuclear well are shown schematically (not to scale for all the figures) for the $D_2$ molecule with a deuteron separation of 0.74 Å, compared with two free deuterons with a classical turning point $r_c \sim 580$ Å for $E_{CM} \sim (1/40)$eV.

Calculating tunneling probabilities for the Coulomb barrier between two d's, and their sensitivity to shielding can quickly make us aware of reasons for pessimism followed by optimism with respect to CF. In the context of $\alpha$-emission, Gamow (1928) first derived the tunneling (transmission) probability $G = e^{-2\Gamma}$ through the mutual Coulomb barrier of two particles of charges $Z_1 e$ and $Z_2 e$, when the center of mass (CM) energy E is much less than the barrier height, in esu

$$\Gamma = \frac{1}{\hbar}\left(\pi Z_1 Z_2 e^2\right)\sqrt{\frac{\mu}{2E}} \qquad (1)$$

where $\mu = m_1 m_2 / (m_1 + m_2)$ is their reduced mass, and $\hbar$ is $(1/2\pi)$Planck's constant. For two d's taking $E \sim (1/40)$ eV for illustration, $G \sim 10^{-2730}$; and in free space the classical distance of closest approach would be ~580 Å. G is extemely small supporting good reason for pessimism about CF, but also illustrating that electron shielding of the barrier cannot be neglected at low E.

There are many models for shielding (screening potential) in a solid which lead to roughly similar results. A model of a spherical shell of radius R of negative charge surrounding each d is the simplest conceptually as well as computationally since it results in only a shifted Coulomb potential

$$V = e^2[(1/r) - (1/R)], \quad r_n \leq r \leq R, \tag{2}$$

where $r_n$ is the nuclear well radius. $G' = e^{-2g}$, where

$$g = \frac{\pi e^2}{\hbar}\left[\frac{\mu R}{2(ER + e^2)}\right]^{1/2} \tag{3}$$

(Rabinowitz, 1990b; Rabinowitz and Worledge, 1989, 1990). For E ~ (1/40) eV and R = 1Å, $G' \sim 10^{-114}$, picking up ~2616 orders of magnitude which illustrates one basis for optimism about CF. In the limit as R $\emptyset$ , g $\emptyset \Gamma$, yielding the unshielded case for $Z_1=Z_2=1$ here, and in general. A model of a uniform cloud of electrons as discussed in Section 4.2.5, as well as use of the Maxwell-Boltzmann (MB) deuteron velocity distribution (Chulick et al, 1990; Rice et al, 1990a,b; Kim 1990b,c,d, 1991a; Kim et al, 1991) would give a significantly higher tunneling and fusion rate but still not enough to account for CF.

Leggett and Baym (1989a,b) presented the most general and most pessimistic limitations concerning the reduction of the d-d Coulomb barrier in a solid by electron screening in the equilibrium state. They concluded that the allowed rate of d tunneling is much too small to account for the inferred fusion rates. They argue that if the effective repulsion of two d's is greatly reduced in Pd and Ti, then these effects should also lead to substantially increased binding of an α−particle to the metal. On the other hand, Azbel (1990) concludes that CF may be possible if there exists a material with a high d concentration, very tightly bound electrons, and a very high metastable state of D. He thinks that this can be met in a solid with narrow energy bands and wide electron gaps, with somehow a proximity of d's ~ 0.1 Å. From equation (3), the tunneling probability for R = 0.1 Å increases to $G' = 10^{-36}$ picking up ~2694 orders of magnitude if a non-equilibrium process could bring two d's as close as 0.1 Å apart in a solid. This illustrates the tremendous sensitivity of G' to the barrier width since a decrease in R from 1 Å to 0.1 Å gains ~78 orders of magnitude in the tunneling probability G'.

Perhaps the most notable theoretical support for CF comes from Julian Schwinger (1990a, b, c) who also contends d's encounter a relatively narrow Coulomb barrier allowing them to fuse into $^3$He in a highly deuterated lattice. He cites Einstein (1907) as pointing out "that the initial phase of a novel investigation can be hindered by an excess of realism". We shall discuss Schwinger's theory in Section 4.2.3.

R.H. Parmenter and Willis E. Lamb (1989,1990) also lend support to the possibility of CF, at the Jones level. The Leggett and Baym (1989a,b) limit is circumvented by both the confining potential for d's in a lattice and a larger effective mass for the conduction electrons at wave numbers the inverse Debye screening length, though they agree that the free electron mass should be used at much larger wavenumbers. (We shall explore the effective mass approach in Section 4.2.5.) The unfavorable milieu for supportive publications is indicated by their comment, "The calculations reported here may be viewed by some as a vain attempt on the part of the authors 'to revive a dead horse,' in view of the recent outpouring of negative publicity concerning cold fusion and the sometimes vicious attacks on its proponents." For them, the low n levels "can be explained without invoking any physics more esoteric than that of screening of positive charges by conduction electrons."

## 3. COLD FUSION PHENOMENOLOGY
### 3.1 General Experimental Results

Hundreds of experimental papers have been published on CF and there have been a number of reviews (Tsarev, 1990,1992a,b; Bockris et al, 1990; Rabinowitz, 1990a; Storms, 1991; Srinivasan, 1991; Preparata, 1991a,b; Tsarev and Worledge, 1991). Our objective is different as it will focus on CF theoretical models, and not the experiments. Nevertheless, for completeness we will briefly cover the main experimental findings with an emphasis on new observations not covered in previous reviews. Many of the claims need further careful verification.

The term "cold fusion (CF)" represents fusion or other anomalous effects at ambient temperature of hydrogen isotopes embedded in a crystal lattice. Observations considered as evidence of CF can be divided into three groups:

A. Data on direct "real-time" detection of products of nuclear fusion. These are mainly neutrons presumably from the reaction (and total energy release Q)

$$d + d \rightarrow {}^3He(0.817 \text{ MeV}) + n(2.452 \text{ MeV}), \quad Q = 3.269 \text{ MeV} \quad (4)$$

where the separate energies are in the CM system. This method is quite sensitive, leading to neutron counts at the rate of $\sim 4 \times 10^{-1}$ to $10^{-3}$/sec. This is the so-called "Jones level" corresponding to a fusion rate $\sim (10^{-22 \pm 1}/\text{sec/dd})(\sim 10^{22} \text{ dd}) \sim 10^{\pm 1}/\text{sec}$. Data on charged particles are less prevalent and more controversial.

B. Accumulation of $^4He$, and tritium (t) products from the reactions

$$d + d \rightarrow p(3.025 \text{ MeV}) + t(1.008 \text{ MeV}), \quad Q = 4.033 \text{ MeV} \quad (5)$$

$$d + d \rightarrow {}^4He(7.6 \text{ keV}) + \gamma(23.7 \text{ MeV}), \quad Q = 23.8 \text{ MeV} \quad (6)$$

in solid electrodes, electrolyte, and evolved gases. Their presence is detected by radiochemistry and mass spectrometry. This method has the advantage of accumulation of these products in a long term experiment. Most groups report an unexpected preponderance of t (or $^4He$) over neutrons. The expected ratios are $t/n \sim 1$ and $^4He/n \sim 10^{-7}$ for reaction (6) (Fowler, 1989). Fowler also estimates that the rate of the reaction

$$d + d \rightarrow {}^4He + e^+ + e^- \quad (7)$$

is $\sim 10^{-2}$ less than reaction (6), so that (7) is not likely to account for the absense of high energy gamma rays in CF experiments.

C. Indications of excess heat release $\Delta Q$ beyond the energy input into the sample when loading it with deuterium. This method is both the least specific as to the nature of the reaction, as well as being the least sensitive. If dd fusion reactions (4) and (5) are responsible for the excess heat, then heat at the 1 W level corresponds to $\sim 10^{12}$ reactions/sec; and for reaction (6) $\sim 10^{14}$ reactions/sec since the $\gamma$'s will likely escape. However, high energy $\gamma$'s have yet to be detected.

Notwithstanding the large diversity of CF experiments all of them can be put into three broad categories:

A. Deuterium Loading (DL) --and/or desorption -- into crystal lattices of various materials such as the transition metals (Pd, Ti, Nb, Zr, etc.) is the primary one. The loading methods vary from electrolysis; gas phase with thermal cycling; ion implantation; and gas discharges. This category of reaction constitutes the main basis upon which CF models are based.

B. Nuclear Mechanofusion (NMF) is fusion accompanying mechanical fracture of solids containing D, which may be related to "fracto-emission" or "mechano-emission" in which electrons and ions are emitted. The experimental situation with NMF is not consistent. NMF can help in the search for new host materials which is not only an interesting problem, but hopefully may lead to greater yields and increased reproducibility. For example, the high temperature superconductor $Y_1Ba_2Cu_3O_7$ was used successfully (Celani et al, 1990) as a host lattice as previously suggested by Rabinowitz (1990a). Positive results on more conventional materials have been reported by the Deryagir group on the mechanical fracture of heavy ice, LiD crystals, and from friction and mechanical activation of Ti surfaces and Ti cavitation in the presence of D (Tsarev and Worledge, 1991). In these experiments, positive results for an excess of neutron emission of 50 to 100% above background have been reported.

C. The results on Nuclear Chemofusion (NCF) were first obtained by a Novosibirsk group (Arzhannikov and Kezerashvili, 1991) that detected neutrons in the course of the chemical reaction $LiD + D_2O \varnothing LiOD + D_2$ and reduction of Pd and Pt from complex salts containing $D_2$. The excess neutron emission relative to the background level was $1.5 \pm .1$. These results were recently supported by the Lebedev-Lugansk group
(Tsarev, 1992b), but still require further verification.

*Critique:* High loading claims (d/Pd > 0.7) must be considered with caution as in some cases this may simply be due to the filling of voids and cracks that are created as the host lattice is forced to expand. Negative NMF results have also been reported (Price, 1990; Sobotka and Winter, 1990; Zelensky and Rybalko, 1991; Watanabe et al, 1992). Using sophisticated statistical software in conjunction with better recording hardware, improved experiments have failed to detect an excess of either neutron or proton emission over background in the case of mechanical fracture of crystals of LiD and $D_2O$ ice. Limits on the NMF rate established by these authors is 15 to 25 times lower than the effect reported by the Deryagin group. Nevertheless, n emission was observed by crushing a lithium niobate single crystal in a $D_2$ gas atmosphere (Shirakawa et al, 1992). Additional critiques follow.

### 3.2 Calorimetry with $D_2O$ Electrolysis

The matter of excess heat release is important for both possible CF practical applications, and possibly the interpretation of the mechanism of this phenomenon. However, the situation with respect to calorimetric measurements remains unclear. The modern CF race started primarily as a result of calorimetric experiments (Fleischmann and Ponns, 1989), followed closely by primarily neutron detection experiments (Jones et al, 1989). Both involved the electrolysis of $D_2O$. Critics pointed out possible sources of error and claimed that the heat release was merely chemical and totally unrelated to nuclear processes. In spite of this, reputable scientists continue to report observations of excess energy release which generally has a sporadic burst-like character, and in some cases is accompanied by particle generation (Storms, 1991; Bockris, 1992; Liaw et al,1991,1993; Bush et al, 1992). The reported heat release $\Delta Q/Q_{in}$ vary from several percent to $\sim 10^3$ percent.

At the Third International Conference on Cold Fusion, held in Nagoya, Japan Oct. 1992, the results of about 30 new calorimetric experiments were presented. Many laboratories (McKubre et al, 1992; Storms, 1992; Kunimatsu et al, 1992) reported that one condition for excess power production is the formation of a highly loaded Pd-D system i.e. $PdD_x$ with x larger than 0.85 - 0.9, taken as an average over the entire cathode. This is clearly seen in Kunimatsu et al , where the excess heat generation becomes prominent around D/Pd 0.85. This was first discussed independentently in 1989 (Pauling, 1989; Golubnichiy et al, 1989a,b; Tsarev, 1992a), but Pauling ascribed a chemical non-fusion character to the effect. Yet, the ~ MJ excess energy over ~ 5 days seen at SRI by McKubre et al does not fit into any conventional chemical explanation. Ota et al (1992) report a possible influence of the mechanical treatment of the Pd cathode.

Especially impressive results were presented by the group from Osaka University (Takahashi et al, 1992). They started their CF experiments in 1989 with very modest results with the observation of very weak neutron emission at ~ 1 n/sec level, and gradually improved their technique and results. During the period of June 1990 to May 1991, their neutron signal increased up to 15 n/sec and they observed two energy components in the neutron spectrum: 2.45 MeV and 3 - 7 MeV. Later using a pulsed current mode of electrolysis operation, they observed t production with t/n ~ $10^5$ - $10^6$ and an excess heat at about 1 W/cm$^3$. [Incidently, the possibility of stimulating CF by various methods including pulses or steps of "current shocks" was first suggested by Golubnichiy et al (1989a,b).] Finally they reached very large, stable and continuous excess heat generation at the 30 - 100 W level. For two months, they registered a net output energy of 160 MJ given an input energy of 410 MJ. The average input power was 50 W; ave. output power 82 W; and ave. excess power 32 W.

*Critique:* Of all the different kinds of measurements that have been made, one might expect the calorimetric measurement to be the most direct and clear-cut as they only involve a temperature measurement, but this is not the case. The possibility of a small systematic error integrated over a long period of time has not been conclusively eliminated. Transient or steady state hot spots due to inhomogeneities rather than CF may give rise to erroneous temperature readings. Changes in the electrolyte may even be a source of energy. It is puzzling that the calorimetric excess energy is ~$10^5$ times higher than the fusion t energy which is ~ $10^8$ times higher than the neutron energy.

Because of the very long charging time in electrolytic CF, it is difficult to prove excess heat generation. A legitimate question may always be raised if the process may be like trickle-charging a battery which after a long period can deliver a lot of power for a short time. If one doesn't keep track, with high accuracy, of the very small energy input during the long charging period, a high energy burst may not really exceed the input. On the other hand if the calorimetry is accurate and the energy is stored in the Pd, it is hard to understand by ordinary storage methods how so much energy density can be stored. Since no known non-nuclear process can deliver > 100 eV/atom, it is much easier to demonstrate that a nuclear process is operative from the calorimetric output energy/atom than from excess heat alone. If the data is reliable, then this has been achieved -- barring one very unlikely possibility. Two caveats must be considered if only energy data (and no nuclear products data) is used. Since all known non-nuclear processes such as chemical, mechanical, etc. only involve the outer electron shells of atoms, they are limited to energies < 10 eV/atom. However, calorimetric data alone would not rule out exotic chemical reactions involving the inner shells of high Z atoms in the host lattice ~ 100eV/atom. As hard as this would be for most chemists to imagine, it is no harder than for most physicists to imagine cold fusion. The second caveat is that even the > 100 eV/atom can disappear if the entire calorimeter is considered rather than just the cathode. Calorimetric measurements are fairly consistent and becoming more convincing.

### 3.3 Detection of $^4$He and Excess Heat Release

Interesting results concerning simultaneous detection of excess heat and production of $^4$He have been reported recently by two American groups who have had their share of critics. Bush et al performed calorimetric measurements and analysis for $^4$He from the gases evolved in the electrolysis of $D_2O$ with LiOD added to the electrolyte (Bush et al, 1991; Miles and Bush, 1992). They found that the evolved gases contained $^4$He in those cases when a considerable heat release was registered. They observed an approximate proportionality between the excess heat $\Delta Q$ and the number density $n_{He}$ of $^4$He. In addition, they found a marked blackening of x-ray films placed in the vicinity of the operating electrolytic cells. A control experiment with the electrolyte replaced with $H_2O$ + LiOH resulted in neither $^4$He nor film blackening.

Based on positive $\Delta Q$ and $^4$He correlation in 8 out of 8 experiments, and lack of $^4$He production in 6 out of 6 control experiments, the authors estimated the probability of a chance coincidence of this result for all 14 measurements as $(0.5)^{14} = 6.1 \times 10^{-5}$. They also claimed to exclude the possibility of atmospheric $^4$He contamination as an explanation of their data. Their estimate for the $^4$He production rate during electrolysis was about $4 \times 10^{11}$ $^4$He/sec/W of input energy. They assert that this quantity of $^4$He exceeds the limit of sensitivity of their method by $> 10^2$. If the process d + d $\varnothing$ $^4$He + $\gamma$ + 23.8 MeV is the source of heat release, then there is an order of magnitude correspondence between $\Delta Q$ and $n_{He}$.

Experiments are reported on the electrolysis of LiD in molten salts of LiCl and KCl at a temperature $> 350$ $^\circ$C with anodes of Pd and Ti, and a cathode of Al at a current density of 700 mA/cm$^2$ by Liaw et al (1991,1993). With Pd they monitored an excess energy release $\Delta Q/Q_{in} \sim$ 600 to 1500%. The maximum excess energy corresponds to 25 W (or 600 W/cm$^3$). The measured excess for Ti was $\Delta Q/Q_{in} \sim$ 100 %. In the same experiment, using mass-spectrometry, they claim a production of $^4$He up to 14 standard deviations above background.

*Critique:* As noted by Tsarev (1992), the presence of large amounts of $^4$He in control samples may make the interpretation of some of these results ambiguous and inconclusive. Nate Hoffman and his Rockwell International colleagues (private communication) confirmed apparent $^4$He production in D$_2$O electrolysis (Bockris et al, 1992). They think $^4$He may be a contaminant in the Karabut et al. (1991) gas discharge CF. They consider the Yamaguchi et al (1992) result an artifact due to $^4$He desorption from glass in this system. They are impressed with but undecided about the Takahashi et al. (1992) results.

### 3.4 Tritium

A number of sensitive methods have been developed for tritium (t) detection that are widely used in CF experiments (Claytor et al, 1991). About 50 papers have reported positive results on anamolous t production during electrolysis. We will only mention 3 papers presented at the 1992 Nagoya Conference.

An original method of CF stimulation was developed at the Los Alamos National Lab (Claytor et al, 1991, 1992) based upon the passage of pulsed current as high as 0.5 A with a voltage up to 3 kV through an alternating system of Pd and Si layers embedded in a D$_2$ atmosphere. Their motivation was that the electrochemical surface layer in an electrolytic cell could be replaced with a metal-insulator-semiconductor (MIS) barrier. Current pulses of 1$\mu$sec to 1 msec duration, with low duty cycle to minimize Joule heating, were passed through the cell to populate non-equilibrium states at the MIS barriers. Where a substantial amount of t was observed, the branching ratio for t/n $\sim$ 3 x 10$^8$ rather than the conventional unity. Kim (1990e) suggested improvements in these type of experiments based on theoretical considerations.

The D$_2$ was high purity, 99.995%, and with very little t contamination. Substantial care was taken that t contamination could not account for the tens of nanocuries produced in successive experiments. Various methods were used for increasing the t production such as increased current density, surface modifiers, and higher D loading. These results are summarized by Claytor et al. All the cells fall into four groups with different effectiveness of t production. Within each group, t production $\propto$ current density. The highest reproducibily achieved t rate was $\sim$ 3.4 x 10$^8$ t/sec. The Luch group in Russia (Romodanov et al, 1992) used a simpler gas discharge technique to saturate various metal samples. A very high t production rate was also observed.

In a recent Texas A &M experiment (Bockris et al, 1992) a standard electrolysis method was used with a Pd cathode and an electrolyte of $D_2O$ + 0.1 Molar LiOD in an open cell. They show the results of t activity in the electrolyte relative to the standard activity in a control cell. The largest t production rate was estimated at $3.8 \times 10^7$ t/sec/cm$^2$. The total number of t produced during 760 hours of electrolysis was reported $\sim 10^{15}$.

Wolf's group first reported large t production (Packham et al, 1989), but later recanted (Wolf et al, 1990) because they considered the t to be present as a contaminant in some pieces of used and unused Pd. On the other hand, Cedzynska et al (1990) report the opposite in finding that t contamination cannot account for the observed t in analyzing 45 samples from a multitude of suppliers including the source used by Wolf.

Srinivasan et al (1990, 1991) report substantial CF t production at the Bhabha Atomic Research Centre (BARC) in India with t/n $\sim 10^8$. The nuclear capability at BARC is world class. Yet, could the t have entered as a contaminant since they have many other t sources there?

*Critique:* Although the above examples of t production to verify CF seem convincing, and illustrate t/n $\gg$ 1, the situation may not be so simple. The use of accumulated t as an undeniable signal of CF has some disadvantages compared with a real time dynamic observation of nuclear products like t, p, and/or n and their energies. No MeV t has yet been observed.

Tritium $\varnothing$ $^3$He+e+$\bar{\nu}$ with a half-life of 12.6 years, and is usually detected by the emitted electron ($\beta$). To detect t, the techniques of $\beta$ activity counting and autoradiography are usually used. The $\beta$'s from some isotopes like $^{210}$Pb look like the beta spectrum from t on a log plot, so a linear plot with an emphasis on the signal peak and end point is needed for the t beta signature to be unambiguous. Mass spectroscopy is used for the detection of $^3$He and $^4$He. Although the sensitivity for t and $^3$He measurements is $\sim 10^{-9}$ lower than that for neutrons, it is possible to integrate over a long time and avoid problems of electromagnetic pickup.

However, it is difficult to exclude spurious effects due to localized t contamination since t is a widespread nuclide which may easily be part of the starting materials. Therefore it is necessary to perform very careful analysis with a statistically significant number of trials from different parts of the sample, and use quality control techniques similar to those developed independently by two Los Alamos groups (Claytor et al, 1991; 1992; Storms and Talcott, 1990).

Tritium may be a contaminant in the electrode materials or in the laboratory surroundings. Chemical reactions can mimic t in the scintillator fluid used to detect t by its β emission. Of course, t can accumulate in the electrolyte as the lighter d's are electrolyzed away in open cell electrolysis. And there is the remote possibility that the electrolyte might be surreptitiously spiked with t. However as shown by Storms and Talcott (Storms, 1991; Storms and Talcott 1990) generation, contamination, and spiking have different easily distinguishable signatures as a function of time. They observed only the generation signal.

Cedzynska and Will (1992) increased confidence in the accumulated t measurements, by developing a procedure for t analysis in Pd with a sensitivity of $5 \times 10^7$ t's corresponding to 1 t per $10^{13}$ Pd atoms for a 0.1 gm Pd sample. They increased the number of tested as-manufactured Pd wire samples to 100, and found no t contamination within the detection limit. Furthermore, Will, Cedzynska et al (1991, 1993) report substantial reproducible t generation in 4 out of 4 tightly closed $D_2SO_4$ cells, with none in 4 $H_2SO_4$ control cells. Although it is unlikely that t entered in as a contaminant, it is important to establish that some other contaminant such as $^{210}Pb$ which mimics the β decay of t did not enter in somehow. This could be done by measuring the half life or $^3He$ production from the t decay.

### 3.5 Direct Detection of Fusion Products

Scores of recent experiments on the direct detection of neutron and proton fusion products were reported at the 1992 Nagoya Conference. Various methods of D loading and desorption were used: electrolysis, gas discharge, thermal cycling, ion implantation, explosive desorption, mechanical fracture, chemical reactions, etc. Most of them reported a statistically signifiant signal to background with the observation of random bursts of n and p. Such observations were first made by Takahashi et al (1990, 1992) with the important finding that there is a high energy component with E ~ 3 - 7 MeV in the n spectrum, in addition to $E_n$ = 2.452 MeV for normal dd fusion. Such components in both the n and p spectra were also noted by a few other groups in the Nagoya Conference.

Previously, many papers reported both positive and negative results on charged particle fusion products such as protons. Cecil et al (1990) reported high power bursts containing up to $10^5$ charged particles. Chambers et al (1990) at the U.S. Naval Research Lab report strong evidence for charged nuclear reaction products with high d ion current densities of 0.4 mA/cm$^2$, and d energies 1 keV.

*Critique:* Direct detection of charged fusion products is a very important source of additional information to supplement the neutron detection data, and in principle could be the most sensitive detection method due to the great sensitivity by which charge can be measured. Unfortunately other than for neutrons, the data on the formation of other nuclear products by electolytic CF may be subjects of controversy. Direct observation of charged particles such as p, t, and $^3He$ is difficult due to their short mean free paths in the solid host lattice, and MeV t have not yet been detected. CF by ion implantation lends itself more readily to such measurements since it allows loading of thin near-surface layers next to a vacuum interface. Thus the charged particles may be measured in vacuum. Even the n"s can be controversial as Cribier et al (1989) point out that a source of n's up to 3.1MeV is $\alpha+d \varnothing \alpha+n+p$ where the α's come from natural occurring radioisotopes. Other α reactions can be a source of n's in light water.

### 3.6 Nuclear Emission and Associated Signal Correlation

Heavy loading with a high concentration of H isotopes results in damage to the host lattice, and for some materials like Pd and Ti can even cause their disintegration. Microcracks, occurring during this process, serve as a source of signals of acoustic and electromagnetic emission. In the context of fracto-acceleration models (FAM), the acceleration of D ions due to the formation of microcracks, results in CF. These models predict a correlation between the production of CF products and emission of acoustic and electromagnetic signals (Golubnichiy et al, 1989a,b).

Such correlations were not firmly established in previous investigations. They were found in some of them (Tsarev, 1990; Tsarev et al, 1991), as well as anti-correlations in others. New extensive correlation data was presented by the Lebedev-Lugansk Russian group (Tsarev, 1992a,b) and the Tokyo Metropolitan Univ. group (Shirakawa et al, 1992). These results appear to support the "fracto-acceleration models" described in Section 4.3.

*Critique:* Strictly speaking, it is possible that the cracking process just accompanies nuclear emission without being causally connected to it. A more detailed study of the time-dependent structure of the signals, and of a three-dimensional localization of their sources would be helpful.

### 3.7 Calorimetry with $H_2O$ Electrolysis

Reports of excess heat production during electrolysis of light water ($H_2O$) were a sensation at the 1992 Nagoya Conference. Claims of such an effect were presented earlier (Mills and Kneizys, 1991; Noninski, 1992) but were very skeptically received. However, as time passed similar results were obtained in five other labs in the USA, Japan, and India. In these experiments, solutions of $K_2CO_3$, $Na_2CO_3$, $Li_2CO_3$, $Na_2SO_4$, $Li_2SO_4$, in $H_2O$ or a mixture of $H_2O$ and $D_2O$ were used as the electrolye. Ni, Ag, Au, and Sn serving as cathodes. In the majority of the combinations of these cathodes and electrolytes, an excess heat production was observed at the level of ~ 3 - 50 % for a rather prolonged period. Srinivasan et al (1992) reported up to 70% excess heat generation above the calibration curve for a few weeks for a mixture of $H_2O$ and $D_2O$ plus $Li_2CO_3$ with a nickel cathode. Many analyzed samples indicated t production even when $D_2O$ wasn't present.

In the experiment by the Hokkaido University group (Notoya et al, 1992; Ohmori and Enyo, 1992) excess heat production in the electrolysis of a $K_2CO_3$ solution with Ni electrodes was found to be correlated with an increase of the calcium ion concentration in the electrolyte. The excess heat is claimed to be consistent with the reactions (Bush and Eagleton, 1992a, b):

$$\begin{aligned}{}^{39}_{19}K + {}^{1}_{1}H &\rightarrow {}^{40}_{20}Ca + 8.33\,\text{MeV} \\ {}^{41}_{19}K + {}^{1}_{1}H &\rightarrow {}^{42}_{20}Ca + 10.3\,\text{MeV}\end{aligned} \qquad (8)$$

*Critique:* These results astounded even CF enthusiasts. Since presently there is no consensus on what constitues "normal" CF, one can hardly call anything abnormal because it all goes counter to the common wisdom. If $H_2O$ fusion is confirmed, the practical significance and nontriviality of CF become manifest. Although the probability for fusing once in the nuclear well is somewhat smaller for p-p than d-d fusion at ambient temperature, the unshielded barrier tunneling probability is some 800 orders of magnitude larger than for d-d fusion just because of the factor of 2 smaller mass. However even though there are some differences with $D_2O$ calorimetry, the $H_2O$ calorimetry seems to represent a kind of discomforting double jeopardy. If it is correct, it is apt to call into question all the experiments in which $H_2O$ was successfully used as a control to bolster confidence in the $D_2O$ results. If it is incorrect, it probably won't just reflect on the particular investigators, but will likely call all CF calorimetry into question. It is important to eliminate $H + H$ and $H_2 + O$ recombination as a possbile source of the exess heat.

### 3.8 Branching Ratio for d + d Fusion

Of particular interest are the results for the t+ p channel of d-d fusion which could prove to be decisive in the selection of CF models -- especially on the crucial question of whether CF only appears cold on a macroscopic scale but is really hot fusion on an atomic scale. In some experiments, an intense production of t has been detected that exceeds the n production by a factor of $10^4$ to $10^8$ (Bockris et al, 1992**;** Srinivasen et al, 1990, 1992; Claytor et al,1991, 1992; Storms and Talcott, 1990; Will, Cedzynska et al, 1991, 1993). If true, this would definitively rule out hot fusion on an atomic scale, but leave as a yet unsolved mystery the cause for such a difference in the n- and t-channels. Conventional experiments with $E > 5$ keV and muon catalyzed fusion experiments with $E \sim$ eV give approximately unity for the branching ratio, though the latter favors the t-channel slightly.

A serious argument put forth against CF is that the 3 MeV p associated with the t channel should also be detectable. Ziegler et al (1989) used Pd foil cathodes as thin as 25 $\mu$m with an adjoining silicon surface barrier (SSB) detector in an attempt to observe the p's. Unlike $\gamma$ and n detectors that have at most 10% counting efficiencies, the SSB has close to 100% detection efficiency for all energetic incident charged particles. They detected no p's above the background level. This may possibly have been because their Pd foil was too thick for 3 MeV p's to traverse if the p's were created at the Pd inner surface rather than in the bulk.

Taniguchi et al (1989) did detect p's by using Pd foil cathodes as thin as 10 $\mu$m, bounded on one side by the $D_2O$ electrolye and by an SSB on the other side. They detected a broad range of proton energies with a cutoff at about 2 MeV. They concluded that the 3.025 MeV fusion p's from the tritium channel lost $\sim 1$ MeV in traversing the full thickness of their Pd foils. This led them to conjecture that the p's were created near the Pd-$D_2O$ interface. As elegant as their experiment is, reproducibility is not yet a routine matter.

*Critique:* Unfortunately as indicated in Sec. 3.4, the situation with respect to t measurements has become rather controversial because of allegations that the observed t signal is a consequence of impurities contained in the starting materials; or spiking. It should be stressed that measurements for t in electrolytic CF experiments are rather complicated and possess multiple stumbling blocks. Therefore, dyanamic gas discharge experiments take on a distinctive significance in that they regularly observe t/n >>1 (Romodanov et al, 1992; Chambers et al, 1990).

The branching ratio is a most important consideration. If real and it is d+d fusion, then CF can't be hot fusion on an atomic scale. What is also needed is a measurement of the p's and their energy. Other than possibly by Taniguchi et al (1989), this has not been done.

## 4. COLD FUSION THEORETICAL MODELS

As yet, there is no consistent theory of CF. So for a systematic discussion of the various models we will choose a classification scheme based upon the different physical assumptions that have been applied. To this end, let us recall that the main difficulty in CF is in surmounting i.e. tunneling through the Coulomb barrier, V(r). At low energies, this barrier prevents nuclei such as d's from approaching each other to a separation r $r_n$ ~ 1 - 5 x $10^{-13}$ cm = nuclear radius, where nuclear forces operate. The probability of tunneling to within this range (Gamow, 1928) for a pure Coulomb barrier, is the Gamow factor given by

$$G \approx \exp\{-\frac{2}{\hbar}\int_{r_n}^{r_c}[2\mu(V(r)-E)]^{1/2}dr\}, \qquad (9)$$

where $r_c$ is the classical turning point for two nuclei with CM energy E, and $\mu$ is their reduced mass.

Traditional fusion occurs inside the sun and stars, and in thermonuclear labs when temperatures, T ~ $10^7$ - $10^8$ K are produced to make E large enough that the integral of equation (9) does not result in an extremely strong suppression of the fusion probability such as at 300 K. We organize our theoretical review into the main categories by which different models choose to overcome this difficulty:

4. 1 Barrier Circumvention (Avoidance)
- *4.1.1 Transmission Resonance (TR)*
- *4.1.2 Lattice Induced Nuclear Chemistry (LINC)*
- *4.1.3 Barrier Free (BF)*
- *4.1.4 Coherent Deuteron Disintegration*
- *4.1.5 QED Neutron Transfer*
- *4.1.6 Bineutron ($^2n$)*
- *4.1.7 Resonance Transparency (RT)*

4.2 Barrier Reduction
- *4.2.1 Heavy Particle Catalylists (HPC)*
- *4.2.2 Superradiance (SR)*
- *4.2.3 Lattice Vibrations (LV)*
- *4.2.4 Quantum Electrodynamic Confinement (QEC)*
- *4.2.5 Screening and Effective Mass*

4.3 Barrier Ascent
- *4.3.1 Fracto Acceleration (FA)*
- *4.3.2 Fracto Acceleration Plasma (FAP)*
- *4.3.3 Interface Acceleration (IA)*

*4.3.4 Lattice Collapse (LC)*
*4.3.5 Quantum Mechanical Transient*
4.4 Narrow Nuclear Resonances (NNR)
4.5 Multibody Fusion
4.6 Exotic Chemistry

We try to confine ourselves in this review to only those CF models which are sufficiently developed in their formalism and quantitative predictions that they are amenable to analysis on the basis of their self-consistency and correspondence with at least some of the experimental data. No model that meets these criteria has been intentionally overlooked. Occasionally, some that do not meet these criteria are also included. Although theory is motivated by the aspiration to explain experimental data, some models properly address only selected experimental results rather than the aggregate possibly due to the uncertainty surrounding the data.

**4. 1 Barrier Circumvention**
*4.1.1 Transmission Resonance (TR)*

The presence of only one barrier leads to a very low transmission probability (coefficient) through it. However anti-intuitively, quantum mechanics allows high probability transit for the one-dimensional problem of particle passage through two (or a periodic sequence) of potential energy barriers for certain discrete values of energy (e.g. at which an odd number of quarter wavelengths fit into the well width).This quantum mechanical effect is due to the dual wave nature of particles. It occurs as a result of constructive interference of waves reflected from multiple barriers (Bohm, 1951), and is called transmission resonance (TR). TR was suggested by Turner (1989) at the Los Alamos National Lab as a possibility for CF, though he apparently did not pursue it further. The idea was further developed by Bush (1990), who assumed that this effect permits the deuterons located periodically in the latttice (such as at interstitial sites) to approach sufficiently close to account for CF.

Bush presumes that a high probability of fusion in a crystal, as compared with nuclear fusion in a free plasma, occurs by a complete passage through the Coulomb barrier at resonant energies of the incident deuterons. Experimental conditions such as temperature and potential are assumed to determine the "resonance" energies of the d's. Changes in these conditions change the fraction of d's getting into the resonant modes. Various dependences of reaction rates are thus predicted as a function of experimental conditions.

To account for the high t/n channeling ratio, Bush speculated regarding polarization of d's colliding at low energies with other d's or nuclei somewhat as in the Oppenheimer-Phillips (1935) model. His view is that upon collision the n is at the leading edge of the d, and the p is at the trailing edge giving reactions which are primarily of the neutron transfer type. Bush accounts for the predominant $d + d \varnothing t + p$ channel by having a d transfer an n to another d to create a t, thus releasing a p in the process. He makes another assumption that in addition to dd fusion, a substantial role is played by TR at the lattice nuclei e.g. $d + {}^{105}Pd \varnothing p(7.35\ MeV) + {}^{106}Pd$, etc. This is how he tries to account for differences between heat release and nuclear products.

*Critique:* There is a basic defect in this TR model. Contrary to Turner, Bush, and Jandel (1991), we feel that Bohm's one-dimensional TR model is not applicable to d's in a lattice, as a given d must also get through the nuclear well of another d. Bohm's model applies to electrons, as they do not have a nuclear interaction. Following Bohm, Jandel has presented his objectons to Bush's TR model of CF, but not to the relevance of the model itself. Even if the model is accepted, one does no better than the conventional extremely low fusion rate. Let us look critically at this model as it is used.

The TR model of CF has a number of inconsistencies beginning with its basic premise. Although the transmission coefficient can be high (but as we shall show not necessarily in the CF milieu), fusion rates can still be extremely low. The build-up of the wave function between the barriers near resonance is a very slow process with time scales ~ the time for alpha decay. As Bohm (1951) points out, the process is similar to the building up of an intense standing wave in a resonant cavity --be it acoustic or electromagnetic. Just as a forced pendulum can acquire a large amplitude over a long time with the application of a small impulse near the resonant frequency, the quantum mechanical wave coming in from the outside acts like the oscillator driver. TR occurs because the wave function is so large in the effective well between the barriers. The unaddressed and unresolved issue in Bush's theory is that it takes too long for the wave function to become large.

The seriousness of the filling time problem can easily be seen quantitatively. For a system of two barriers, with each described by the potential $V(x)$, the transmission probability (coefficient) as obtained from the WKB approximation is:

$$P' \quad [1 + 4G^{-4} \sin^2\{(\pi - J)/2\}]^{-1}, \tag{10}$$

where $G = \exp(-J)$ is the Gamow factor given by equation (9).

Let us consider two d's approaching each other through two barriers at room temperature, with $E = kT = 0.025$ eV. Taking into account screening of the Coulomb potential barriers, $G \sim 10^{-100}$ as shown in Sec. 2. If the system is far from the resonant energy $E_{res}$, $P' \sim G^4 \sim 10^{-400}$. This corresponds to a small fraction of $\sim (G^2)(G^2)$ of the incident d's tunneling through two barriers. At resonance $J = \pi$, and equation (10) gives a unity tunneling probability, i.e. $P' = 1$ for any G. The resonance is related to the existence of a metastable state whose lifetime is $\Delta t \sim \hbar / \Delta E$, where the half-width $\Delta E = E - E_{res}$ for $P' = 1/2$. Bohm shows that the lifetime is

$$\Delta t = t_t G^{-2}, \tag{11}$$

where $t_t$ is the classical transit time to cross the well and return. For a well width of $\sim 1\text{Å} = 10^{-8}$ cm, and a velocity $\sim 2 \times 10^5$ cm/sec, $t_t \sim 10^{-13}$ sec. Thus at resonance, $\Delta t \sim 10^{-13}$ sec $(10^{200}) = 10^{187}$ sec. The age of the universe is small in comparison, being only $\sim 15 \times 10^9$ years $= 4.7 \times 10^{17}$ sec. Of course shorter times are possible as E gets further from resonance, but the combination of lifetime and tunneling probability does not appear capable of accounting for CF.

Bush (1990) has tried to eliminate the problem of the metastable state in TR by introducing a "stagnant wave" associated with a "chain" of barriers, but this approach appears futile.

### *4.1.2 Lattice Induced Nuclear Chemistry (LINC)*

The main theme of a series of papers by Chubb and Chubb (1990a,b) relates to the wave nature of boson particles in a solid. They feel that just as electrons are better described as waves in a solid, deuterons should not be described as particles when they are inside a solid lattice. The many-particle wave function for $N_d$ deuterons (bosons) can be represented in the form

$$\Psi(\mathbf{r}) = (N_d!)^{-1/2} \sum_{\mathbf{r}_m} \prod_{m=1}^{N_d} \psi_{Bloch}(\kappa_m, \mathbf{r}_m) \qquad (12)$$

$$\psi(\mathbf{k},\mathbf{r})\exp[-\varepsilon(\mathbf{k})t/h] = N_L^{-1/2} \sum_{n=1}^{N_L} \phi_n(\mathbf{r},t)\exp(i\mathbf{k}\cdot\mathbf{R}_n) \qquad (13)$$

where $\phi_n(\mathbf{r},t) = \phi(\mathbf{r}-\mathbf{R}_n,t)$. $\mathbf{R}_n$ are the coordinates of potential wells which the d's are in. $N_L$ is the number of host unit cells. Summing up is performed over all exchange of coordinates $\{r_i\}$.

This wave function includes terms corresponding to n-tuple occupation of a well by the deuterons since they are indistinguishable bosons. For example, for two d's in two wells:

$$\psi \sim [\phi_1(\mathbf{r}_1,t)\exp(i\mathbf{k}_1\mathbf{R}_1)+\phi_2(\mathbf{r}_1,t)\exp(i\mathbf{k}_1\mathbf{R}_2)][\phi_1(\mathbf{r}_2,t)\exp(i\mathbf{k}_2\mathbf{R}_1)+\phi_2(\mathbf{r}_2,t)\exp(i\mathbf{k}_2\mathbf{R}_2)]$$
$$\sim \{\phi_1(\mathbf{r}_1,t)\phi_1(\mathbf{r}_2,t)\exp(i[\mathbf{k}_1+\mathbf{k}_2]\mathbf{R}_1) + \phi_2(\mathbf{r}_1,t)\phi_2(\mathbf{r}_2,t)\}\exp(i[\mathbf{k}_1+\mathbf{k}_2]\mathbf{R}_2). \qquad (14)$$

The presence of such terms is a critical point for the authors which gives them reason to state, "Overlap of the wave functions necessary to initiate the reaction is ensured by algebraic properties of a many-particle wave function, but not by tunneling which is the basis of conventional nuclear physics."

The basic predictions of the LINC model are a high rate of fusion in a lattice, dominance of the production of $^4$He, and heat release without observable fast nuclear products.

*Critique:* The authors' premise identifying simple overlap of the d wave function with fusion in a lattice is erroneous. This is because their wave function is derived from a Hamiltonian which neglects the d-d interaction, and does not minimize the full Hamiltonian of the d system in a lattice. If the dd Coulomb repulsion is taken into consideration, then the zeroth-order approximation results in a wave function in which the very terms needed for LINC of the type $\phi_m(\mathbf{r}_1,t)\ldots\phi_m(\mathbf{r}_n,t)$ are missing. For example, for two d's a correct zeroth-order approximation wave function takes the form:

$$\psi \sim \phi_1(\mathbf{r}_1,t)\phi_2(\mathbf{r}_2,t) + \phi_1(\mathbf{r}_2,t)\phi_2(\mathbf{r}_1,t) \qquad (15)$$

This is a well known rearrangement of the wave function in the zeroth-order approximation when the interaction is accounted for.

Their excess energy release is due to a total neglect of dd Coulomb repulsion, which if included would give a tremendously smaller fusion rate.

### *4.1.3 Barrier Free (BF)*

Perfect crystals do not exist above cryogenic temperatures because the minimization of the free energy implies that a crystal must have entropy associated with lattice imperfections. Vysotskii et al (1990) make a virtue of what otherwise may be considered a vice. They take lattice imperfections in the form of voids with characteristic dimensions ~ 10 Å, in which they quantize the motion of the fusing particles with individual wave functions $\psi_n(r)$ and energy levels $E_n$. A distinction is made between D's which are fermions, and $D_2$'s and d's which are bosons. [Since it is not crucial, for brevity we shall not carefully distinguish between them.]

Due to a sign change of the potential for some optimal void sizes ($R_{opt} \spadesuit$ 7.5 Å for a spherical void), the energy of the dd interaction $\langle V_{dd}(r_1 - r_2) \rangle$, averaged over spatial and spin variables, is found by the authors to be much less than $\Delta E = E_{n+1} - E_n$. And the non-diagonal matix elements of this interaction are considered to be as small. In their opinion, it therefore follows that in spite of the strong Coulomb interaction, the motion of each d for n>>1 is independent and is characterized by the same single-particle wave function $\psi_n(r)$ as in the case of the missing atoms had they still been in the void.

In a particular embodiment of their model, a spherical infinite square well potential has $N \sim 10^3$ d's placed in it. Accordingly, the main quantum number for most of the atoms is great, i.e. n >> 1. Therefore $|\psi_n(r)|^2 \sim 1/V'$, where V' is the volume of the void. Thus for them in the absence of the Coulomb repulsion, the probability of close approach (fusion) for two d's is determined not by tunneling through a barrier $|\psi_{dd}(0 < r < r_{nuc})|^2$, but just by the well (void) volume:

$$\int_V |\psi_1|^2 |\psi_2|^2 dr \sim \frac{1}{V'} \quad . \tag{16}$$

The smaller the void volume, the higher the probability, provided the volume is big enough to accommodate D's. The main consequence of this model is the artificial disappearance of the Coulomb barrier in the mocrovoid. This leads to an enhancement of the fusion rate by a factor of $10^{30}$ higher than for a $D_2$ molecule.

*Critique:* Their idea is a little like the free electron model of a metal in which the Coulomb interactions between electrons can be neglected to an excellent approximation because of the Pauli principle and the compensating Coulomb background of the ions. This permits one to obtain wave functions and energy levels without the complication of introducing the Coulomb potential into the Hamiltonian for the system. This is correct for the depiction of a metal, as close electron encounters requiring high energy are not paramount to the properties of a metal. However they throw out the very interaction of relevance for CF.

Their main inconsistency is the incorrect consideration of the dominant term in the potential energy in the dd interaction Hamiltonian, as a small perturbation fcompared with the kinetic energy. Indeed, let us consider a three-dimensional infinite square well of radius R into which are placed N atoms of deuterium (D) interacting with the potential $V(r_i - r_j)$. Neglecting $V(r_i - r_j)$, the kinetic energy of the atoms at the highest level is the Fermi energy for this system,

$$E_F \approx \frac{h^2}{21m_D} \frac{N^{2/3}}{R^2} \sim 10^{-3} eV, \qquad (17)$$

for $N = 10^3$, and $R = 10^{-7}$ cm. The neglected mean Coulomb potential energy is •V® = $e^2/r$ ~ 20 eV, since the average particle spacing r ~ $R/N^{1/3}$ ~ $10^{-8}$ cm. Therefore the potential cannot be treated as a small perturbation in comparison with the kinetic energy. On the contrary, consequently due to the large D mass, $m_D$, the highest kinetic energy $E_F \ll$ •V®.

The ground state is the minimum total energy state of the system. Since $E_F \ll$ •V®, it is also  the minimum potential energy state. Thus for a system of number density $\eta$ as constrained by the well dimensions, the average particle spacing $r = \eta^{-1/3}$ ~ $R/N^{1/3}$. Note that for high quantum numbers (n >>1), the case considered by them is met by the condition $\kappa R \gg 1$, where $\kappa$ ~ n/R is the wave number related to the D motion. Consequently, only at kinetic energies   10 MeV may one speak of perturbation theory with regard to the $V_{DD}$ potential. However, no void in a solid body is capable of confining deuterons of such high energy.

### *4.1.4 Coherent Deuteron Disintegration*

Hagelstein (1990,1991,1992,1993) presents three variations of his model of CF as a coherence of virtual neutron, n*, transfer (emission and absorption) and n* states in a solid heavily loaded with d.

#### *A. Coherent depp Reaction*

He assumed that virtual n emission in the solid state may be due to inverse β-decay of the d: d + e ∅ n* + n* + ν, which is the depp reaction of the weak interaction capture of an electron resulting in the emission of a neutrino (v) and two off-shell (virtual) neutrons. For low "subthreshold" eV energies, n* can be captured by another "external" d only within ~ $10^{-13}$ cm. An esimate of the cross section for this process is not presented. It is claimed that the n* emission rate ∝ $N_D^2$ = (the number of d's)$^2$.

Hagelstein (1990) says, "The depp scenario ... possesses a number of recognized serious weaknesses...." He feels that the "coherent neutron transfer reactions .... are so superior to the depp scenario as to have made the old scenario completely obsolete."

*Critique:* The author's papers present equations related to quantum theory in general, and in particular solid state physics and the single mode laser without quantitative estimates as yet. One general problem with his approach is that because of the large n mass, it cannot exist very long as a virtual particle n*. This greatly limits the n* range to ~ $10^{-13}$ cm, and the number of d's with which it can interact because of the extemely small volume of interation $10^{-39}$ cm$^3$.

Even for a high d density ~ $10^{22}$ d/cm$^3$ this puts an upper limit of $10^{-17}$ reactions/n*. The virtual neutrons exist only an extremely small fraction of the time. Hence if quantitative estimates were to be made of the fusion rate, it would be too low.

*B. Neutron Transfer due to Macroscopic Magnetic Field*

An external macroscopic magnetic field of flux density B ~ 1 Gauss is assumed to be able to enhance d fusion in a solid (Hagelstein, 1990, 1991). It is claimed that the interaction $\mu_n \cdot B$ with the neutron anomalous magnetic moment $\mu_n$ leads to n transfer from "donor" nuclei in one cell to "acceptor" nuclei in another cell. The energy released in this reaction is presumed to be absorbed by phonons and transformed into heat.

*Critique:* The general formulae presented seem to be unrelated to the claim of coherent n transfer from one cell to another. In the author's formulae (Hagelstein, 1990), the perturbation manifests itself in the value of the matrix element for the transition from the initial d state to the intermediate state of the "disintegrated" d:

$$\int \phi_n^*(r_n) \phi_p^*(r_p) [\mu_n \cdot B] \phi_D(r_n, r_p) dr_n dr_p . \qquad (18)$$

This value is close to zero if one takes into account quasihomogeneity of the external field and orthonormality of the intermediate states. Thus it is our opinion that the magnetic field perturbation cannot influence the steady state dd fusion rate in a solid.

*C. Coherent Neutron Transfer Dynamics in a Lattice*

Assuming that the problem of n transfer from one cell to another is solved, he treats the d system in a lattice by analogy with a laser. The role of population inversion is played by the number of neutrons, $N_n$. Hagelstein (1990 estimates the n transfer reaction rate to be

$$\left[\frac{d}{dt} N_n^*\right]_D \propto \frac{N_n N_D \mu_n B}{\hbar} \left[\frac{V_n}{V_L}\right] \sim 2 \times 10^7 \, B / \sec, \qquad (19)$$

where B is the macroscopic flux density in Gauss. For him, $N_n = 10^3$, the number of D's is $N_D = 10^{23}$, the neutron magnetic moment $\mu_n = 3 \times 10^{-12}$ eV/Gauss, and the ratio of nuclear to lattice volumes is $V_n/V_L = 10^{-40}$.

*Critique:* Unfortunately equation (19) relating to the virtual neutron production rate is not the CF rate, and the CF rate is not estimated. It is not clear to us that his description is quantitative in predicting excess heat and t production. The barrier is avoided with the reaction mechanism involving weakly interacting neutrinos and two off-shell neutrons so there is no Coulomb replusion. However the price for this is an exremely low reaction rate.

*4.1.5 QED Neutron Transfer*

Danos and Belyaev (1991) also try to explain CF on the basis of n transfer, with quantitative predictions. For them it is in the context of perturbation theory in the second order approximation for the electrodynamical interaction of the dd system with an extra "external" nucleus C': d+d+C'$\varnothing$p+t+C'. A first deuteron, $d_1$ is in the bound state $|O\rangle$ with a binding energy ~ 0.1 eV. The Coulomb field of the nucleus C perturbs the d wave function, so that the transition amplitude into the intermediate state $\langle n|V|O\rangle$ is different than zero for the highly excited levels $|n\rangle$. This is in particular for the states with the disintegrated d with the following Feynman diagram: A n produced by the d disintegration is captured by the second free $d_2$. The excess momentum q is transferred to the nucleus C' in the final state.

They calculated the amplitude corresponding to this diagram by the standard method, $U_{mo} = \sum_n \langle m|V|n\rangle\langle n|V|O\rangle / (E_n - E_o)$. This leads them to estimate the fusion rate ~ $\alpha 10^{-5}$/sec for a beam of $d_2$'s moving with velocity v ~$10^5$ cm/sec. Here the factor $\alpha$ ~ $10^{-4}$ - $10^{-8}$. Comparing their fusion rate with the conventionally expected rate, they optimistically conclude that the ~$e^4$ term in quantum electrodynamics (QED) is ~ $10^{40}$ - $10^{50}$ times larger than the direct term. Thus they conclude that CF is consistent with conventional physics and does not require an exotic solution.

*Critique:* This model is a variation of a previous QED model for CF (Danos, 1990). Formally the calculation of the above Feynman diagram contribution does give a high fusion rate. However, for the case being considered, of nonrelativistic d's, the Born approximation cannot be applied to describe their Coulomb interaction $Z_d Ze^2/r$, $Z_d = 1$, with the nucleus C'. The parameter $Ze^2/hv$ in this case >> 1, whereas $Ze^2/hv$ << 1 is required for the Born approximation to be valid.

The physical shortcoming of their derivation is related to the following point. In calculating the matrix element $\int \varphi_n^*(r)(Ze^2/r)\varphi_o / r \, dr \sim \langle n|V|O\rangle$ it is necessary to take into account the strong perturbation of the nonrelativistic state $\varphi_o(r)$ caused by the repulsive Coulomb field of the nucleus C'. As a result, the correct wave function $\varphi_o(r)$ near r ~ 0 is $\propto$ the exponentially small Gamow factor G responsible for the extremely small probability for a d approaching the nucleus C'. Further, in order to produce the intermediate state $\varphi_n(r)$ with d disintegration, a large nuclear momentum q ~ $[m_D(1 \text{ MeV})]^{1/2}$, must be transferred which is much larger than the characteristic momentum of the bound state $\varphi_o(r)$.

This means that one needs to calculate the asymptote in $|q|$ for $\int \exp(iqr)(Ze^2/r)\varphi_o / r \, dr$ which is determined by the singularity of the product $\varphi_o(r)V_{Coul}(r)$ at r = 0. Since $\varphi_o(r\varnothing 0)$ is exponentially small, the matrix element will be similarly small. The same reasoning is applicable to vertex 2. One can say roughly that instead of the Coulomb factor $1/q^2$ in their Feynman diagram, it is necessary to use the factor $G/q^2$.

Inapplicability of the Born approximation in this problem can be seen from the following simple example. It is well known that the first Born approximation always gives the correct result for the Coulomb scattering cross-section. However, it is inapplicable for the calculation of the nonrelativistic wave function near r ~ 0 for a repulsive potential since

$$\frac{|\phi_{\text{correct}}(0)|^2}{|\phi_{\text{Born}}(0)|^2} \approx \frac{2\pi e^2}{\hbar v} \exp\left[-\frac{2\pi e^2}{\hbar v}\right] \sim G \ll 1. \tag{20}$$

### 4.1.6 Bineutron ($^2n$)

This model of the formation of two coupled n's, a bineutron ($^2n$) under e- capture is similar to that of Section 4.1.5: $e + d \varnothing\ ^2n + \nu_e$. It was proposed independently by Andermann (1990), Pokropivny et al (1990), and Russell (1990,1991a,1991b), and later by Yang (1991). The $^2n$ energy levels were estimated from pn scattering data. The energy of the singlet $^2n$ state was calculated to be ~ 125-143 keV by Pokropivny et al. This is close to that for d, ~ 67 keV. The $^2n$ lifetime was estimated by different approaches as ~ $10^{-15}$ sec. They claim this is enough for CF. They assume that the lifetime may be increased up to between $10^{-9}$ and $10^{-7}$ sec, if the electron momentum can be transferred to a group of atoms or the entire crystal as in the Mossbauer effect.

The authors think that this model can explain, n, t, and He production and their absence in some experiments, as well as the sporadic nature of CF. They think that an electron with the needed energy ~ 60 keV may be produced in the electrolytic double layer, during fracture, etc.

*Critique:* There are two main objections against the bineutron model. First, the experimental data from nuclear physics testifies against the existence of $^2n$. Second, the probability for the weak interaction electron capture reaction is extremely small. Even if one overlooks the necessity of having electrons with very high energy to initiate such a reaction, the small probability makes this mechanism incapable of explaining CF.

### 4.1.7 Resonance Transparency (RT)

Kim et al (1993) have proposed a resonance transparency (RT) solution of CF based on a new generalized Coulomb barrier transmission coefficient which simultaneoulsly accommodates both non-resonance and transmission resonance contributions in the tunneling. The resonance behavior results from an interplay of the transmitted and reflected waves involving both Coulomb barrier and nuclear interaction, and is distinguished from conventional resonances such as narrow neutron capture resonances, which are primarily due to nuclear interaction. For zero angular momentum, they find the transmission coefficient $T_{KZ}(E) = \dfrac{4s_o \overline{K}_1 R}{|(\Delta_o + is_o) - (\overline{K}_2 R - i\overline{K}_1 R)|^2}$, which has a Breit-Wigner form when $(\Delta_o + is_o) - (\overline{K}_2 R - i\overline{K}_1 R) = 0$ at a pole in the complex E plane. Even though the resonance is extremely narrow, they propose that energy sweeping in a non-equilibrium process(Kim, 1991b) gives high yields. This could explain why system perturbations seem to be required for CF.

*Critique:* Similar objections may be raised here as for Section 4.1.1 regarding filling time. However, they may not apply since there is only one barrier.

## 4.2 Barrier Reduction
### *4.2.1 Heavy Particle Catalylists (HPC)*

The genesis (Frank, 1947) of these models is muon catalyzed fusion (MCF). Here the electron in a deuterium atom is replaced by a muon. Since the orbital radius is $\propto 1/(\text{mass})$, the muon orbit in a D is 207 times smaller than the electron orbit. This allows d's to come much closer to the attractive nuclear force in a muonic $D_2^+$ molecule with only a thin barrier to tunnel through as shown schematically in Fig.2. MCF was quickly pursued theoretically by Sakharov (1948), experimentally observed by Alvarez et al (1957), and analyzed by Jackson (1957) who calculated the high rate of $\sim 10^{11}/\text{sec}$ for the muonic d-d reaction. MCF may properly be called cold fusion since it occurs at ordinary temperatures, though it appears to have little in common with the CF we are considering. Moir et al (1989) suggested that MCF resulting from cosmic ray muons might explain CF. However, they subsequently withdrew this proposition (private communication; cf. Sec. 4.2.5).

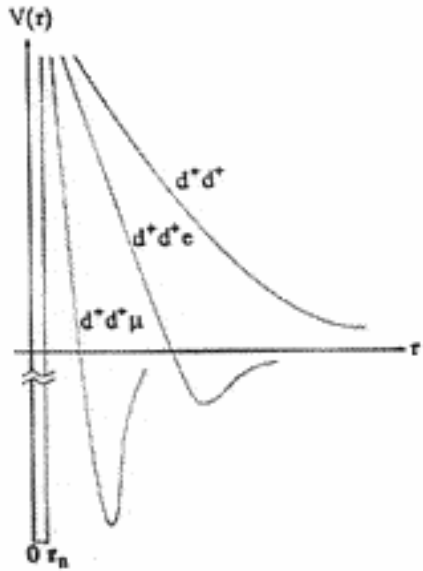

**Figure 2.** Coulomb barriers are compared schematically for two free deuterons, a $D_2^+$ ion with deuteron separation of 1.11 Å = 1.11 x $10^{-8}$ cm, and a muonic $D_2^+$ ion with deuteron separation ~ 540 fm = 5.4 x $10^{-11}$ cm. The barrier reduction qualitatively exemplifies that of the models of Section 4.2 such as the Lattice Vibrations of Schwinger, Muon Catalyzed Fusion, Heavy Particle Catalysts, Superradiance, etc.

Heavy particle catalysis (HPC) is a variation of MCF. It relates to as yet undiscovered particles (X) which possesss a number of unusual properties. Candidates to play the role of such particles come from some modern physics models of the elementary particles. They may have been formed in the early stages of the Universe. Or they may be created at extremely high energies by cosmic ray interaction. In either case, their accumulation on the Earth is possible in principle. Unlike muons, these particle catalysts are supposed to be stable (or long-lived), heavy, and subject to nuclear repulsion at small distances. Different versions of HPC use different particle catalysts:

1) Shaw et al (1989) postulate free diquarks with ~ 1 GeV mass, and 4/3 electron charge.

2) Bazhutov et al (1989) posit an isodoublet of neutral and charged hadrons with 100 GeV masses. The hadrons contain heavy u-quarks, $X^o = Uu$, $X^- = Ud$, and complexes of the type $Y_N = X^o n + X^- p$, and $Y_d = X^o d$.

3) Rafelski et al (1990) propose super heavy negatively charged hadrons.

4) Edward Teller suggested an undiscovered particle which he jokingly dubbed a "meshouganon" (daft) particle (private communication).

Ordinarily the CF scenario suggests that X particles are adsorbed on the surface of some nuclei in the matter surrounding us in space. They are delivered to the Earth in the form of complexes of the Xp type. It is expected that in a deuterium-rich medium the most probable type of nuclear interaction is pd exchange: $Xp + d \varnothing Xd + p$. It is suggested that the uncorrelated emission of neutrons in CF experiments is the result of catalysis by a flux of Xp constantly coming down from outer space on the Earth's surface with a density ~1 cm$^{-2}$/sec.

The events of a burst-like nature are associated with the release of X particles (that were adsorbed on ordinary matter nuclei) under some non-equilibrium conditions. For example in Bazhutov (1990) they arise from the FAM (fracto-acceleration mechanism). Following this, the X particles are capable of interacting in various catalysis stages, e.g.:

$$d + d + X \rightarrow \begin{cases} {}^3He + n + X \\ T + p + X \\ {}^4He + X \end{cases} \qquad (21)$$

Shaw et al (1989) infer that for comparitively light X particles, the two particle channel of the reactions (21) dominates by a factor of $10^3$ to $10^6$ times higher than the three body channels, from considerations related to the short-range repulsion and phase space. The reaction rate via this two body channel may be very high, ~ $10^{10}$ sec$^{-1}$X$^{-1}$, due in particular to a charged state Xd. However according to Rafelski et al (1990), for superheavy X particles, the relationship of the channels is just the opposite of this.

Since the neutron binding energy in nuclei exceeds that in the d by 5 MeV, the Xd system can give a neutron to any of the surrounding nuclei, which leads to excess heat and to energetic protons:
$$Xd + A \varnothing X + p + A'. \qquad (22)$$

The general consequences of different versions of MHPC are as follows:
a) A high fusion rate due to the strong screening of the Coulomb potential.

b) Bursts as a result of high rates of "cycling" ending with the possible emergence of the X particle from the CF cell, or just sticking to a heavy nucleus in the cell.

c) Poor reproducibility as a consequence of the low concentration of the X particles; their escape or capture; sample preparation procedure; their occurance as a function of the depth of the mineral from which the cathode material was extracted; etc.

d) Large freedom regarding the properties of the catalysts allows for $t/n \gg 1$. There are also some differences in the predictions.

*Critique:* The HPC models are based on the theoretical hypothesis of the existence of heavy stable particles. This is one of a number of different hypotheses that are being considered in modern high energy physics and cosmology for various reasons. One reason is to account for the missing dark matter of the universe. We consider that although this hypothesis is extremely speculative, in principle it can serve as a testing ground to try to unify widely disparate fields of physics. On the other hand, it may be quite premature to explain the enigmatic mechanism of CF -- which phenomenon is itself in dispute -- with the help of still more exotic phenomena. Many of the properties of such particles are not established, and this leaves a wide horizon of arbitrariness for their authors.

Of course if such particles are experimentally detected, or if CF matures to the point that strict critical tests of HPC can be conducted, then the final judge can be experiment, as it should be. However, present attempts to detail such models and give quantitative predictions have met with failure. As was shown by Zamiralov et al (1993), predictions of the model of Bazhutov et al (1990) turned out to be completely unjustified.

*4.2.2 Superradiance (SR)*

Bressani et al (1989), and Preparata (1990a,b,1991a,b) propose that the key to understanding CF lies in superemissive dynamics -- superradiance (SR) -- in a solid. According to these authors, this means that the components of elementary atomic systems to some extent lose their individuality and become part of a kind of collective plasma. This plasma is a medium of charged particles vibrating about their equilibrium positions with plasma frequencies $\omega_p = e\sqrt{n'/m\varepsilon}$, where e and m are the charge and mass of the particles, n' is their number density, and $\varepsilon$ is the permittivity of the medium. They have an instability in the quantum electrodynamic (QED) ground state (independent zero-point oscillations). Their minimal energy state is a superradiant one in which all the plasma particles oscillate in phase with the electromagnetic field that is excited coherently from the perturbative ground state of QED. A three-dimensional modulation of this field has typical dimensions $\lambda = 2\pi c/\omega_p$, in a region where the motion of matter and field is coherent.

SR was first presented by Dicke (1954), though his work is not referenced in any of these papers. Coherence is the key to understanding SR. If we have N particles radiating incoherently, then the total power radiated is the sum of the individual powers. For particles radiating coherently (in phase) in a small enough volume that the phase coherence is not lost from one end to another, then the total electric field $E_t$ is the sum of the electric fields in each of the radiated waves. The total radiated power $\propto E_t^2$. For example, in simple terms, if the individual radiated power is P for each particle, then $P_{incoherent} = NP$, and

$$P_{coherent} \sim N^2 P. \tag{23}$$

This implies greatly enhanced radiation for a large number of particles since $P_{coherent} \sim NP_{incoherent}$.

On the basis of their SR dynamics calculations, their conclusion is that SR provides a very strong effective screening of the Coulomb dd potential by the electron plasma in Pd. At high concentrations $C'' = d/Pd \sim 1$, the d's are expected to form a coherent region in a lattice which interacts as a unified quantum-mechanical system. The energy released in nuclear processes in the coherent regions is surmised to be transferred to the electron plasma, and is thus transferred to the lattice as heat. This results in very little energy transfer to the d plasma, which is a requirement if coherence is to be maintained.

From the viewpoint of tunneling, enhanced screening is equivalent to increasing the d energy to $E_d \sim 100$ eV. It is claimed that this alone increases the CF rate up to the Jones level $\Lambda_{dd} \sim 10^{-21}$ to $10^{-20}$/sec /dd without even consideration of coherence. In this incoherent regime, the relationship between the dd fusion channels is expected to be about the same as in an ordinary plasma with $t/n \sim 1$.

At high concentrations with $C'' \sim 1$, owing to the contribution of coherent interactions, the reaction rate is supposed to increase $\sim 10^{10}$ times higher than in the incoherent regime, corresponding to CF at the Fleischmann-Pons level. Here, a direct energy transfer to the lattice explains the heat release which exceeds the nuclear product flux (assuming t is at the low n level i.e. $t/n \sim 1$) $\sim 10^{10}$ to $10^{12}$ times. Another configuration, as related to the Oppenheimer-Phillips (1935) mechanism allows $t/n \gg 1$.

*Critique:* In addition to conceptual errors, there are a number of analytic and numeric errors which make the agreement with CF results much more problematic than the author's claims (Preparata,1991a). He makes a $> 10^2$ numeric error. His $r_n = 20$ fm is too large a value for the d-d nuclear attraction distance. The deuteron-deuteron nuclear force diameter is < 8 fm, resulting in a the tunneling probability $D_T < 10^{-42}$ rather than his $10^{-40}$ due to the great sensitivity of $D_T$ on $r_n$. This $10^2$ error increases considerablly due to a serious shielding error. At the top of his p.88 rather than "an enhancement of some 30 orders of magnitude over the tunneling amplitude for molecular deuterium", the resulting number should be much less. Consequently, not even the Jones level of CF is obtained. His equation (26) is crucially important, but is not derived in detail in this nor any other of their papers. It appears overly simple for the complex Feynman diagram from which it is claimed to result.

The use of the SR approach may be considered a highly imaginative attempt at reformulating old established theoretical field concepts with possibly new twists. The main object of this work contradicts quantum mechanics in its overestimates of superscreening to achieve CF, as we shall next demonstrate.

For a screening estimate, an assertion is made that Z electrons orbiting in phase about a Pd nucleus look like a sphere of radius $\delta$ defined by the dispersion of the plasma oscillations

$$\delta = \left[\frac{\hbar}{2m_e\omega_p}\right]^{1/2} \approx \frac{6.7 \times 10^{-9}}{Z^{1/4}} \text{cm}. \tag{24}$$

Due to electron oscillations, they assume a d may be covered by a cloud of Z electrons. According to these authors, the screening potential of these electrons takes the form

$$V_{screen} \equiv V_s \approx -\frac{Ze^2r^2}{2\delta^3} \text{ for } r \leq \delta. \tag{25}$$

At $r = r_o = \delta(2/Z)^{1/3}$, $V_s + V_d = 0$, since $V_d = \frac{e^2}{\delta}$ at $r = \delta$. Observe however that with this abnormally large amount of screening the tunneling probability $P' = G^2$ is exceedingly high especially for $E_d \sim 0$:

$$P' = G^2 \approx \exp\left[-\frac{2\sqrt{2\mu}}{\hbar}\int_{r_{nuc}}^{r_o}\left(\frac{e^2}{r} - \frac{Ze^2r^2}{2\delta^3}\right)dr\right] \approx 10^{-42}. \tag{26}$$

To understand this anomalous result, note that $\delta$ is close to the characteristic dimension of a Thomas-Fermi atom of atomic number Z,

$$R_{TF} = \frac{0.885\hbar^2}{me^2Z^{1/3}} = 4.5 \times 10^{-9} \text{cm}/Z^{1/3}. \tag{27}$$

Consequently, the arrangement of Z electrons inside the region r   $r_o \sim R_{TF}$ is possible only in the Coulomb potential Ze/r of a nucleus of charge Z, but not in the field of a singly charged d. Physically, Z electrons cannot take part in the steady state screening of the d Coulomb field, as a common effective potential for the dd interaction. Solids would collapse if such close equilibrium screening were possible.

### 4.2.3 Lattice Vibrations (LV)

According to Schwinger (1990a,b,c) the effective potential of the d+d and also p+d interactions are modified due to averaging related to their zero-point oscillations in a solid lattice. In simpler words, the coupled harmonic motion of particles is supposed to lead to a reduction of the Coulomb barrier for fusion. When calculating this effect, one replaces the coordinate **r** in the Coulomb potential e/r by **r** + δ**r**, where the operator addition δ**r** corresponds to d oscillations and has a conventional expansion in terms of phonon degrees of freedom

$$\delta\mathbf{r} = \bullet\sum_q \sqrt{\hbar/2m_d\omega_q N}(\mathbf{a}_q e^{-i\omega_q t} + \mathbf{a}_q^+ e^{i\omega_q t}). \tag{28}$$

Here $\mathbf{a}_q$ is the boson operator for phonon production with q momentum; N is the number of d's, $\hbar\omega_q$ is the energy of the q th phonon. When averaged to first approximation in the ground state, the effective potential of interaction for a slowly moving proton with a phonon oscillated d is

$$\langle 0 | V(\mathbf{r} + \delta \mathbf{r}) | 0 \rangle \sim \begin{cases} \dfrac{e^2}{r} & r \gg \Lambda \\ \dfrac{e^2}{\Lambda} & r \ll \Lambda \end{cases} \quad (29)$$

Where $\Lambda = \sqrt{\dfrac{\hbar}{2m_d}\left\langle\dfrac{1}{\omega_q}\right\rangle}$, and $\left\langle\dfrac{1}{\omega_q}\right\rangle$ is an average in terms of phonon modes. For $\langle 1/(\hbar\omega_q)\rangle \sim 1/(0.1\text{eV})$, we obtain $\Lambda \sim 10^{-9}$ cm. In the next approximation

$$\delta V \approx -[\nabla V(r)]^2 \bullet_q \left[\dfrac{1}{2m_d N}\left\langle\dfrac{1}{\omega_q^2}\right\rangle\right] \quad (30)$$

Assuming $\left\langle\dfrac{1}{\omega_q^2}\right\rangle \approx \left\langle\dfrac{1}{\omega_q}\right\rangle^2$ and considering that $2\Lambda \sim 10^3 \hbar^2/(m_d e^2)$, Schwinger finds that

$$\delta V = -\dfrac{e^2}{r}\left|\dfrac{10\,\Lambda}{r}\right|^3 . \quad (31)$$

This expression is valid for $r \gg \Lambda$. For $r \gtrsim 10\,\Lambda$, eq. (31) gives a rather marked decrease of the Coulomb potential, $e^2/r$. Schwinger concludes from this that a substantial suppression of the Coulomb barrier may be possible at the expense of lattice vibrations (LV).

*Critique:* Our analysis indicates that there is a limit to what the phonons can do, and that there is not a sufficiently strong effect from LV. To us, Schwinger's LV approach appears applicable only for $\delta V \ll V$. Accordingly, this implies a small relative correction to the repulsive Coulomb potential rather than the large one he finds. As we showed in Section 2, modest decreases in the width of the Coulomb barrier can have enormous increases in the tunneling probability P' (the more so the lower P' is to begin with). However, the LV decrease in V goes further than seems warranted, and gives too large of an increase in P'. A separate point is that in going from the d+d reaction (1990a) to the p+d reaction (1990b,c), the 23.8 MeV γ is avoided and only a 5.5 MeV γ has to be absorbed directly by the lattice. What is the upper energy limit of such a process? We hope that Schwinger will address the issues raised and clarify the situation.

There are also two minor issues that may need resolution for the d+d case. 1) The competing decay channels d+d $\varnothing$ t+p and d+d $\varnothing$ $^3$He+n normally occur in times $\sim 10^{-22}$sec. The maximum frequency of the phonons $\sim 10^{13}$/sec implying a phonon emission time $\sim 10^{-13}$sec. It is not obvious *a'priori* that these decay channels will not depopulate the d+d scattering state faster than the phonon emission. 2) Schwinger (1990c) suggests that his model can produce t production rates $\sim 10^3$/sec $-10^{10}$/sec. It is not obvious that even with dE/dx energy degradation of $\sim$ MeV t due to electron and ion interactions, that the reaction

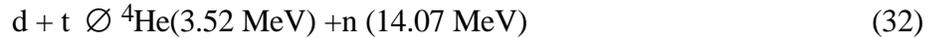

$$d + t \; \varnothing \; {}^4\text{He}(3.52 \text{ MeV}) + n\ (14.07 \text{ MeV}) \quad (32)$$

cannot take place in measurable quantities. Reaction (38) has not been observed in CF. This issue also applies to other models and experiments which have a large t production.

-29-In agreement with us, detailed calculatons by Szalewicz et al (1989) and Petrillo et al (1989) of the effects of d oscillations in a lattice greatly limit the LV effect. They derive an upper limit increase in P' < a factor of $10^8$, which is quite insufficient to attain even the Jones level. Crawford (1992) is even more pessimistic, claiming that "properly treating deuteron motions would lead to smaller calculated p-d fusion than if phonons were neglected."

### *4.2.4 Quantum Electrodynamic Confinement (QEC)*

To explain CF, Jandel (1990) assumed the existence of a hypothetical confinement phase with strong coupling in quantum electrodynamics. Earlier (Celenza, 1986 et al; Fukuda, 1989) this phase was introduced for other purposes in connection with interpreting narrow $e^+e^-$ resonances found in collisions of heavy ions. It is conjectured that QEC can arise in very strong electromagnetic fields analogous to color confinement in quantum chromodynamics. The electrons in QEC are always bound with positrons, forming neutral electromesons; or with atomic nuclei, forming neutral electronuclei. A d in the confinement phase is supposed to be coupled with an electron, forming a neutral electrodeuteron $D^*$ with strong coupling. Equivalent to ordinary experience, two $D^*$ can form a $D^*$-$D^*$ molecule; and He nuclei can form $^3He^*$ and $^4He^*$ states.

Nuclear dd fusion in QEC proceeds either by way of direct tunneling in the collsion of two $D^*$, or via a two-step process in which $D^*$-$D^*$ is first formed, and then a nuclear reaction is thought to follow inside this bag. CF in a deuteride is sustained at the expense of absorption of d's into a region in the QEC phase from the surrounding deuteride. The size of this region grows until various processes violate the conditions for the existence of this phase. Jandel thinks that the QEC regions can occur spontaneously, or be induced artificially as for example by irradiation with heavy ions or in the decay of transuranium nuclei.

With dimensions of a $D^*$-$D^*$ complex ~ $10^2$ x $10^{-13}$ cm, it is possible to explain the high energy releases of the Fleischmann-Ponns type. If chemical equilibrium is achieved in the QEC region, the yield of various reaction products is defined by their masses and by the condition of isospin conservation. Thus the main product is $^4$He. In the initial period of the QEC regime until the attainment of chemical balance, an abundant formation of t is possible. Further predictions depend on additional assumptions, and are not specific to the given model.

*Critique:* It should be stressed that QEC is based on an assumption related to quantum electrodynamic phase properties that does not strictly follow from first principles. Although confinement conditions for the strong coupling constant in high energy physics have been demonstrated in QED for calculations on a lattice, this has not been established for coupling at the much lower level of the fine structure constant. Thus far QEC provides no real testable predictions as the known experimental data is itself used as input parameters (calibration) for the theory. It would be more realistic if the model calibration were based on input parameters outside the field of CF, such as from the collision of heavy ions.

### *4.2.5 Screening and Effective Mass*

Some of the earliest papers suggested more effective shielding (screening) of the Coulomb barrier in a lattice than in a free space plasma as the explanation for CF. The effects of screening and high densities on fusion rates were considered long ago in other contexts. The concept "pycnonuclear" was introduced (Cameron, 1959) for fusion reactions that depend mainly on density. Harrison (1964) coined the term "cryonuclear" for fusion reactions that are temperature independent and occur only in degenerate nuclear gases, as well as discussing many screening options. Though he was concerned with much higher densities, his analysis is relevant to CF. He even speculates on the possibility of superfluidity for a boson gas ($D_2$'s and/or d's ) at high density. He thus anticipated a number of present-day hypotheses regarding screening and CF.

The simplest approach to shielding, as described in Section 2, was presented by Rabinowitz (1990b) with a shifted Coulomb potential given by equation (2) which permits an analytic solution to the Schroedinger equaton. Another approach recognizes that in a metallic lattice, there are no forces between d's at r = R, a position of stable equilibrium resulting in the shielded potential
$$V = e^2[(1/r) - (3/2r) + (r^2/eR^3)]. \tag{39}$$
Each d polarizes the lattice electrons to about the Debye length. Charge neutrality implies a uniform spherical cloud of one electronic charge for equation (39). Assuming uniform charge density underestimates the shielding, as the positive nuclear charge will tend to increase the electron charge density as r decreases.

As a d moves through a metallic lattice, it attracts the nearly free valence electrons which screen its field. Takahasi (1968, 1971, 1990, 1991) argued that a non-linear dynamic response produces a four-times greater electron density. Combined screening by the electrons and d's is more effective than that due to electrons alone. Whaley (1990) makes a case that because of coherent effects, d screening may exceed electron screening. Rambaut, (1992) argues for a large number of shielding electrons in double screening. Burrows (1989) applies only Thomas-Fermi electron shielding.

Muon catalyzed fusion (MCF in Secs. 2 and 4.2.1) is a direct form of strong screening. However, Cohen and Davies (1989) argue that MCF is a most improbable solution for CF since the muon stopping flux is not high, the fraction captured by D's is low, and the fraction retained by D's is even lower. Because the muonic molecule formation is slow, and because a significant fraction stick to He at fusion, each muon can catalyse only a few reactions before decaying in about 2.16 μsec. They conclude that even under the most favorable conditions imaginable, MCF could not begin to account for the n production rate of 0.41/sec [$4.1 \times 10^{-3}$/sec measured with an efficiency of 1%.] reported by Jones et al (1989).

Koonin and Nauenberg (1989) find that an electron mass increase by factors of 5 to 10 would be needed to account for CF in $H_2$ isotopes. For the free electron mass, they find that the rate of d+d molecule fusion is $3 \times 10^{-64}$/sec/dd, some 10 orders of magnitude greater than previous calculations. They also find that the rate of p+d molecule fusion is ~ $10^{-55}$/sec/pd. This is much higher than the d + d rate because of enhanced tunneling in this lighter system. However, even though the lighter p+p molecule system has a higher tunneling rate, its fusion rate is the lowest because of the much lower astrophysical constant S(E), the fusion probability after the nucleons tunnel into the well. So conventionally, light water reactions seem very unlikely.

An interesting feature of some models is the supposition of quantum mechanical effects in the host crystal lattice leading to a high electron effective mass with a substantial increase in the fusion rate. Thus a central role is played by the periodic character of the inter-atomic fields in crystals (periodic potential). However Garwin (1989) argues that a large electron effective mass (inversely proportional to the second derivative of the energy dispersion with respect to the wave vector in the band structure of a lattice) is not capable of allowing 2 d's to approach closer than an effective mass equal to the free mass. Rabinowitz and Worledge (1989, 1990) suggested that the periodic potential can lead to a reduced effective d mass in a solid which may account for CF. As pointed out in Section 2, Parmenter and Lamb (1990) make a strong argument that a larger effective mass for the conduction electons at wave numbers the inverse Debye screening length accounts for CF, at the Jones level. They exceed the Leggett and Baym (L and B) maximum limit of $3 \times 10^{-47}$/ (sec-dd) by $10^{24}$, where $10^{17}$ comes from the confining potential (1989) for d's in a lattice (neglected by L and B), and another factor of $10^7$ comes from the effective mass of the conduction electrons.

*Critique:* Very close equilibrium screening is not possible, or solids would collapse. Nevertheless, this arena has strong pro and con arguments, making it difficult to judge. It is important to determine whether the pro positions of Parmenter and Lamb, and others like Azbel's (1990) really get around the L and B (1989a, 1989b) limitations as discussed in Section 2. It is hoped that further dialogue can take place on this important issue. It is important to bear in mind that the L and B argument against the effectiveness of shielding is an equilibrium argument, and that CF is not necessarily an equilibrium process. Although Garwin (1989) is negative on shielding, the role of effective mass, and CF in general, he doesn't close the door entirely by saying,"experiments will surely show whether cold nuclear fusion is taking place; if so,it will teach us much besides humility and may indeed provide insight into significant geophysical puzzles."

**4.3 Barrier Ascent**

Dynamical models (DM) to explain cold fusion are in the category of being the most intuitive and the furthest developed. They are based on the idea of overcoming the Coulomb barrier at the expense of energy that has been acquired by d's in various processes either in the lattice or on the surface of the solid. Thus DM
are basically hot or at least lukewarm fusion models with characteristic d energies $> 10^2$ eV. They most commonly achieve d acceleration by means of electric fields. Figure 3 schematically shows the reduction in barrier width with increased energy. Experimentally it may be possible to distinguish between cold fusion and hot fusion on an atomic scale (which is the basis of some CF models) by temperature and kinematic broadening and shifting of the characteristic fusion product lines (Kim et al 1992). We were drawn to the solar neutrino problem in the hope that insight into this hot fusion paradox might also help us understand CF. We think our findings help to solve this problem(Kim et al 1993a,b,c).But,it does not appear helpful for CF.

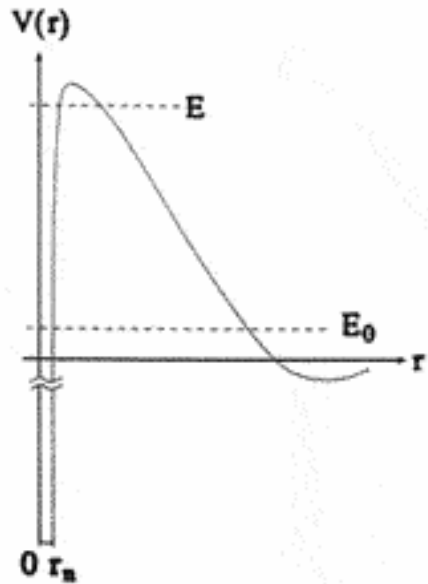

**Figure 3.** The narrower Coulomb barrier thickness due to barrier ascent by increased energy E is depicted as related to the dynamical models of Section 4.3 such as Fracto Acceleration, Interface Acceleration, Lattice Collapse, and Quantum Mechanical Transient.

*4.3.1 Fracto Acceleration (FA)*

This version of the DM implies that the accelerating electric fields arise as a consequence of the appearance of electric charges on the newly formed surfaces of cracks. These cracks develop either due to a direct externally imposed mechanical effect on a d containing material, or as a result of hydrogen embittlement. In fracto-emission, the emission of electrons, ions, and photons (triboluminesence) are experimentally well known (Dickenson et al, 1990; Tsarev, 1990),though not so well understood theoretically. Hydrogen embittlement occurs when hydrogen or its isotopes are loaded heavily into the crystal lattices of various materials to the point where materials like Pd and Ti simply disintegrate. Both phenomena, hydrogen embrittlement and fracto-emission are at the heart of the FA explanation of CF. Reports related to FA are that electrolytic cathodes exhibiting anomalous nuclear signals have their surfaces covered with a network of microcracks; whereas speicimens exhibiting no activity were relatively free of microcracks. FA was first suggested (Kluev et al, 1986) to explain Nuclear Mechanofusion, and only later (Golubnichiy et al, 1989a,b) for CF, during D loading and deloading as reviewed by Tsarev (1990), and Chechin et al (1990a,b).

Order of magnitude estimates for the FA fusion rate can be obtained with the use of a simple model of the accelerating crack as a plane parallel capacitor (Tsarev, 1990; Chechin et al, 1990a,b). In the context of FA, the possibility exists of relating the Jones level of neutron emission rate with the amplitudes of accompanying electromagnetic and acoustic signals, as produced by crack propagation. Golubnichiy et al (1989b) considers these points in detail.

As a consequence of the possible relationship between CF and hydrogen loading fracture, FA leads to the following predictions:

1) The statistical nature or randomness of CF is a consequence of the stochastic nature of the fracture process.

2) There is a possible surface to volume relation of CF since FA activity is manifested in a layer of hydride subjected to fracture in a given time interval. Progressively new layers of the sample participate in the process as fracture proceeds into the sample depth.

3) A possible quasiperiodicity in the bursts is a consequence of the dyanamics of growth diffusion of a hydride layer down to a certain critical thickness L, at which point it fractures. Characteristic time intervals between bursts are $\tau \sim L^2/D_o \sim 10^2$ to $10^4$ sec, where $D_o$ is the diffusion constant for the process.

4) A necessary condition for FA is a non-equilibrium state in the metial-hydride system in which there is a fracture of the hydride, and the creation of an unstable hydride phase.

5) Externally imposed influences can affect fracture, and hence possibly CF rates.

6) There are changes in the physical-chemical properties such as resistivity, and structure during periods of activity.

7) Correlations may exist between the emission of nuclear synthesis products, and electromagnetic and acoustic signals which are known to accompany crack propagation.

8) Since FA is essentially hot fusion on a microscopic scales, the relative fusion rates inherent to hot fusion should be observed in CF. The ratio of these rates should be $\Lambda(p+d): \Lambda(d+d): \Lambda(d+t) \sim 10^{-5}: 1: 10^2$.

*Critique:* To say the least, it has been difficult to obtain confirmation of the straightforward predicition 8) of the relative fusion rates. This difficulty is only in part due to dE/dx loss of energy in the solid. The FA model also has several other problems. First of all, to quantitatively explain the experiments on CF, one has to assume the presence of rather stong fields in the microcracks $\sim 10^7$ to $10^8$ V/cm. These are extraordinarily high, and for steady state would lead to electrical breakdown with the field dropping precipitously. However, such high transient fields may be possible on a very short time scale. The relationship between the various time scales characterizing both cooperative and competitive processes such as crack propagation, establishment of the electric field, acceleration time, field collapse, discharge time, and field emission time has not been sufficiently addressed.

Evidently, it is necessary to assume an apparently arbitrary enhancement in the dielectric properties of the transition metal hydrides during the period of nuclear activity. This is necessary to inhibit electrical breakdown (discharges) and field emission, since both processes rapidly degrade the electric field. Arguments in favor of this assumption (Golubnichiy et al, 1989a; Tsarev, 1990), relate to two well-known solid state mechanisms. One of them is a consequence of the band structure of solids which results in an enhancment of the dielectic properties with increasing interatomic distances as occurs with the expansion of hydride lattices with increased hydrogen loading. Another is removal of electrons from the conduction zone, which also increases the insulating properties. A well-known example is the rare earth metal hydrides (Tsarev, 1990).

Preparata (1990b) tries to ensure a large enough time for the existence of a large charge density on the sides of a crack in terms of his superemissive (superradiance, SR) model for the solid. This is an application of his conjecture of Sec. 4.2.2 to FA, since he has the cracks filled with coherent radiation generated in the solid by plasma oscillations of the nuclei. For an ordinary plasma, the maximum ponderomotive potential created by this field, and acting on a particle with charge e and mass m is $V \sim n' m_{Pd} d_o^2 \left(\frac{e^2}{m}\right)$ where $d_o$ is the amplitude of oscillation of the plasma particles. As an example, let m be the electron mass, $m_{Pd}$ is the mass of a Pd atom, the number density $n' \sim 10^{23}/cm^3$, and $d_o \sim 10^{-9}$ cm; then $V \sim 160$ keV. For a deuteron, $V \sim 40$ eV. This sub- stantially increases the dd fusion rate compared with a thermal energy of 1/40 eV.

In a physical sense, the quantity -V plays the role of a negative work function. We see that he gets too large a magnitude for V. Evidently this is due to an overestimate of the amplitude of the plasma oscillations as a free parameter. In addition, according to his scheme highly energetic electrons ~ 100 keV should be emitted from such a process, which no one has detected.

### *4.3.2 Fracto Acceleration Plasma (FAP)*

The particular model considered here is another possible acceleration mechanism in cracks or voids. As with the other DM, it depends on breakage of the hydride material with the formation of charge (due to electrostatic bond breaking) on the sides of the newly formed void or crack which can provide accelerating potentials (Tsarev, 1990). The FA models of Sec. 4.3.1 neglect the role of gas in the voids. Zelensky (1990) has given an argument with quantitative estimates to support the DM by including the effects of both gas loading and electrical gas discharges in the voids. A similar mechanism was noted by Golubnichiy et al (1989a,b).

They reason that if the crystal lattice surrounding a void is free of imperfections, pressures as high as $10^5$ atmospheres may be attained in the process of deuteration at heavy loading. Zelenski argues that if this is accompanied by the accumulation of charge inside the void, this can lead to an ionization rate $\Theta$ of between $\sim 10^{-10}$ to $\sim 10^{-2}$ of the D in the void leading to very high voltage differences. Thus for example, for a uniformly charged sphere of radius $R = 10^{-2}$ cm, filled with D at a pressure $P = 10$ atm, and $\Theta = 1.7 \times 10^{-6}$, the difference in potential energy between the center and the surface is

$$\Delta V = e\rho_e R^2/6\varepsilon \sim 1.5 \times 10^4 \text{ V} \tag{34}$$

where e is the electronic charge, $\rho_e$ is the charge density, and $\varepsilon$ is the permittivity. The electrostatic repulsion pressure $P_{elec} = \frac{1}{2}\varepsilon E^2$ < P, and as a rule should not prevent charge accumulation within the context of this model.

*Critique:* One weakness in the FAP model, which also applies to the other acceleration models, is the ion acceleration mechanism. In view of the fact that at such high pressures the mean free path $\lambda$ of the ions is so short, it is not clear how the ions can attain the full potential difference. At 10 atm pressure, $\lambda \sim 10^{-6}$ cm, and the average energy gain may be limited to

$$E_G \sim \frac{\lambda}{R} \Delta V \sim 1.5 \text{ eV}. \tag{35}$$

Even if some ions attain $10^2 E_G$, this is only $\sim 150$ eV. Since $\lambda \propto$ (number density of D)$^{-1} \propto P^{-1}$ for constant temperature, the ions can attain a higher fraction of $\Delta V$ at lower pressure. However, at lower pressure the possible $\Delta V$ is also much lower due to lower $\rho_e$, charge leakage, lower breakdown voltage, etc.

On the other hand, Zelenski (1990) points out that even higher energies than given by eq. (34) may be possible due to a strong collective entrainment interaction in a spark breakdown between an electron beam and the plasma ions so that the energy of the d's may get up to

$$E_d \sim (m_d/m_e) E_e \gg E_e. \tag{36}$$

This equation was also derived by Rabinowitz and Worledge (1990) as will be discussed in Section 4.3.3.

The FAP model also has many of the problems noted in Sec. 4.3.1. The FAP at least comes to grips with the question of the characteristic time periods. This is a critical issue for: the crack propagation time which is assumed to be the same as the charge accumulation time, $t_c$; the charge relaxation (discharge-emission) time, $t_r$; and the acceleration time, $t_a$:

$$t_c \sim 10^{-7} \text{ to} 10^{-8} \text{ sec}; \; t_r \sim 10^{-6} \text{ to} 10^{-9} \text{ sec}; \; t_a \sim \frac{10^{-12}}{\sqrt{P\Theta}} \sim 10^{-9} \text{ sec} \tag{37}$$

### 4.3.3 Interface Acceleration (IA)

Rabinowitz and Worledge (1990) suggested the following interface acceleration (IA) model based upon an observation of dendrites on electrolytic cathodes. Asperities (sharp microscopic whiskers) grow on electrolytic cathodes (Lin et al, 1990), and as a way of relieving internal and external stresses in a variety of settings (Rabinowitz and Garwin, 1968). Field enhancement ~height/(tip radius) at the tip of the whisker, (Rabinowitz 1968) together with the already present high double-layer electric field, can lead to very high local electric fields $\sim 10^7$ V/cm even though the macroscopic field is very low. A high current density of electrons can be field emitted from a whisker inside a low pressure bubble at the interface between the $D_2O$ and the cathode, and ionize some D's.

These authors pointed out that if a small number of d's become entrained with the high current density of electrons, the d's would attain the same velocity v as the electrons -- much the same as a log in a river attains the velocity of the current in the river. The ratio of the energy of the d's to the energy of the electrons would be the same as their mass ratio

$$\frac{\frac{1}{2}m_d v^2}{\frac{1}{2}m_e v^2} = \frac{m_d}{m_e} = 3670. \tag{38}$$

Even though the potential difference in a $D_2O$ bubble is only a small fraction of the voltage applied to the cell, they pointed out possible non-equilibrium mechanisms for producing larger transient voltages. Thus for an electron energy ~ 10 eV, the d energy could be as high as 37 keV. Some consequences of the IA model for CF are discussed by Kim (1990c,d; 1991a).

Predictions of the IA model are:

1) Fusion rates at the Jones level and higher.

2) Sporadic character and burst-like nature of the process as whiskers are damaged.

3) Incubation period related to the growth of whiskers.

4) Poor reproducibility of the data related to the irregular behavior of the whisker growth process, and of the occurrence of non-equilibrium discharge conditions.

*Critique:* Unlike the fracto-acceleration models, no relation between CF and cracking is implied in the IA model. This is not supported by those experiments that claim a correlation between CF products and acoustic signals. However, although whiskers can grow due to stress during temperature cycling and rapid phase transitons in Ti, there is no applied electric field in the cycling experiments. Perhaps electron acceleration with entrainment of the d's could be related to field enhancement by whiskers of fracture produced electric fields.

In the context of CF experiments produced by gas discharges, IA may also be a possible explanation. Such experiments (Karabut et al, 1991) do have an incubation period which could be related to whisker growth.

The same serious criticism that applies to the other acceleration models applies to IA. Namely, that the branching ratio, and rates for the different reactions should be the same as for hot fusion, but appears not to be so.

### 4.3.4 Lattice Collapse (LC)

Rapid phase transitons in Ti in the temperature cycling experiments may be an important part of the CF mechanism. In this connection, lattice collapse (LC) may apply to both these experiments as well as others such as the electrolytic experiments. The LC model of Tabet and Tenenbaum (1990a,b,c) implies that the release of energy stored in a metal-deuteride ($MD_x$) lattice and its transfer to d's may be related to the transition of the $MD_x$ system from a homogeneous phase of concentration $C_O$ to a heterogeneous one where the d's are distributed in individual regions with a high concentration $C_+$ and a low concentration $C_-$.

The initial transition point is above the critical temperature $T_c$ for phase transition where the D atoms are uniformly distributed inside the metallic lattice. A more complex phase exists below $T_c$ where the three different regions can emerge relative to the D/M loading. Since the lattice constant is concentration dependent, the three phases are characterized by different elastic energy. Supposedly, when the concentration decreases from the uniform value $C_O$ to $C_-$, the excess elastic energy is released abruptly by the collapse of micro-regions.

They calculate the elastic energy stored in a micro-region of radius R in a face-centered cubic lattice as

$$E_R = 8\pi v^2 m \left(\frac{R}{a_o}\right)^3 \epsilon^2, \tag{39}$$

where, $\epsilon = (a_o - a_-)/a_o = 0.2(C_o - C_-)/3$. Here m is the metallic atomic mass; v is the average sound velocity, $a_o$ and $a_-$ are the lattice constants corresponding to the concentrations $C_O$ and $C_-$. It is assumed that when a critical radius $R \approx a_o/\epsilon$ is reached, the stressed region propagates out to such distances at which the atomic displacement reaches the equilibrium lattice constant $a_o$. Then the stressed region suddenly collapses down.

Collapsing occurs when the elastically stored energy in the stressed region proves to be the energy required for the displacement of the external layer to a new equilibrium position. This is a little like ordinary spallation effects in that a wave is produced that converges toward the center of the region. This displacement wave drags (entrains) the d's along at high velocity causing some of them to collide with each other in the interstitial spaces near the center of the collapsing region. The LA model predicts a burst-like nature for CF as determined by the non-equilibrium conditions that lead to phase transitions. They estimate that for $E_R \approx 160$ eV, optimal values of $C_O \approx 0.25$ to $0.35$ produce a fusion yield ~ $10^2$ to $10^5$ n per gram-atom of Pd.

*Critique:* One criticism of the LC model is that it transfers too much elastic energy to a pair of d's in the center of the collapsing region. It is likely that the energy transfer efficiency is low, and that much of the energy reflects back into the lattice in the form of a rarefaction wave. In terms of channeling ratio and yields for various fusion reactions, the LA model is subject to the same criticisms as the other acceleration models since it is basically a microscopic hot fusion model.

### *4.3.5 Quantum Mechanical Transient*

Chechin and Tsarev (1992, 1993) propose enhancement of the barrier penetrability caused by the generation of high momentum components in the d wave function due to the introduction of transients. They consider a bound s-wave state of $\psi(r,t)$ of two d's with a spherically symmetric potential $V_o(r)$ in a crystalline lattice. If the Hamiltonian of the system instantly changes at $t = 0$ to another potential $V(r)$ that does not lead to bound states, then the total fusion probability is

$$W = \left(\frac{A}{4\pi}\right) \cdot \left(\frac{2M}{\hbar}\right) \cdot Q, \tag{40}$$

where $M = m_d/2$, $A \sim 1.6 \times 10^{-16} cm^3/sec$, and $Q = \int_0^\infty |\tilde{\chi}_o(k)|^2 k \, dk / |F(k)|^2$. Here F(k) is a Jost function for the Hamiltonian H with the potential V(r), where

$$\tilde{\chi}_o(k) = \int_0^\infty x_o(r) \psi_k(r) dr.$$ $\psi_k(r)$ is the complete system of eigenstates of the Hamiltonian H: $H\psi_k = \kappa^2 \psi_k$.

This allows us to understand the origin of the possible fusion probability enhancement in the model. Conventionally, the main fusion suppression factor is due to the small probability of finding two d's in the region $r \quad r_{nuc} \sim 10^{-13}$ cm. This fact corresponds to the exponentially large Jost function at $k \sim k_{at} \sim m_e e^2/\hbar^2$. However in the Fourier transform of the initial state $\tilde{\chi}_o(k)$ there are also large $k >> k_{at}$ for which $|F(k)| \approx 1$. Correspondingly, the integral of k is determined by the high momentum tail of the amplitude $\tilde{\chi}_o(k)$. This can considerably affect the fusion probability.

It was supposed that such transients in the potential might arise in some processes in solids such as on the surfaces of hydride cracks (Tsarev, 1992b). The total number density of microcracks after a long hydration of Pd might reach $n_{cr} \sim 10^9 - 10^{12}/cm^3$ leading to reasonable CF rates.

*Critique.* For systems like a $D_2$ molecule with relatively low rigidity and localization, the role of transient induced fusion appears negligible compared with conventional steady state fusion. So there is no strong reason to expect the solid state to change this.

### *4.4 Narrow Nuclear Resonances (NNR)*

The theoretical fusion cross section is

$$\sigma(E) = [S(E)/E]P', \tag{41}$$

where E is the energy of the fusing particles in the CM system, and P' is the tunneling probability for penetrating the barrier. S(E) is the astrophysical function obtained experimentally for energy measurements above ~ 10 keV, and extrapolated to lower energies (Chulick et al, 1993). S(E) is a measure of the probability that the particles will fuse after penetrating the barrier. Up to this point, the CF models have avoided or attacked the barrier in one way or another. Another possibility for increasing the fusion rate at low E, is to have a larger than expected S due to a resonance.

Enhancement of the fusion rate as a consequence of a possible narrow resonance of a nuclear origin such a ($^4$He)* resonance in the vicinity of the threshold of its decay into two d's were suggesed (Shihab-Eldin et al, 1989; Kim, 1990a). A considerable increase in the cross-section of some fusion reactions as a result of the existence of narrow resonances at low energies is well known. It plays, in particular, an important role in astrophysical nucleosynthesis.

If such resonances have a very small width and are near the dd fusion threshold, they would not have been seen in scattering experiments, but might nevertheless strongly influence fusion at very low energies. The main prediction of this model is an increase in the CF rate resulting from an as yet undiscovered long-lived resonance.

*Critique:* The available data on cross-sections of the dd and dt reactions seem sufficient to most thermonuclear physicists to get a reliable extrapolation of the astrophysical S-factor to low energies. No such resonance has yet been found, and there is no guarantee that it would be large enough to account for CF. How- ever, as Chulick et al (1993) have shown, extrapolation of the S factor is not a closed issue; and there is a precedent for the $3\alpha$ resonance in $^{12}C$ that permits nucleosynthesis in red giant stars to get past $^{8}Be$.

**4.5 Multibody Fusion**

Four independent conjectures (Becker, 1989; Rabinowitz, 1990a; Kim, 1990c; and Takahashi, 1991) have been made suggesting the possibility of multibody fusion. Rabinowitz pointed out that although the probability for a three-body collision in free space is extremely smaller than a two-body collision -- in a solid, "Channeling increases the probability of a nearly one-dimensional collision, with essentially the absence of angular momentum in the final state. This may permit low energy resonances which greatly increase the fusion cross-section -- particularly for energy and momentum conserving three-body collisions." These authors then compile many body reactions. We list some here and their total energy release Q, as they may be of interest. The separate energies are in the CM system.

$$d + d + d \varnothing \; d(15.9 \text{ MeV}) + {}^4He(7.9 \text{ MeV}), \quad Q = 23.8 \text{ MeV} \quad (42)$$

$$d + d + d \varnothing \; d(4.75 \text{ MeV}) + {}^3He(4.75 \text{ MeV}), \quad Q = 9.5 \text{ MeV} \quad (43)$$

$$d+d+d+d\varnothing \; {}^4He(23.8 \text{ MeV}) + {}^4He(23.8 \text{MeV}), \quad Q = 47.6 \text{ MeV} \quad (44)$$

$$p + d + p \varnothing \; p(4.12 \text{ MeV}) + {}^3He(1.37 \text{ MeV}), \quad Q = 5.5 \text{ MeV} \quad (45)$$

$$d + p + d \varnothing \; p(19.0 \text{ MeV}) + {}^4He(4.76 \text{ MeV}), \quad Q = 23.8 \text{ MeV} \quad (46)$$

$$d+d+p+p\varnothing \; {}^3He(5.5 \text{ MeV}) + {}^3He(5.5 \text{MeV}), \quad Q = 11.0 \text{ MeV} \quad (47)$$

$$p+d+d+d\varnothing \; {}^3He(16.8 \text{ MeV}) + {}^4He(12.6 \text{MeV}), \quad Q = 29.3 \text{ MeV} \quad (48)$$

There are two main consequences of this hypothesis. 1) In multibody fusion reactions, the released energy is carried by some particles that cannot escape the solid. So this energy heats up the lattice without the emission of visible nuclear products. 2). In reaction (46), energetic protons are released. In reactions (42,43,44, 46, and 48), energetic alphas ($^4He$) are released which in turn can produce energetic n's and p's via the process $^4He + d \varnothing \; {}^4He + p + n$. This might explain the high energy components that have been seen (Chambers et al, 1990; Cecil et al, 1990; Takahashi et al,1990, 1992). Takahashi (1991) has suggested a number of multibody fusion scenarios in a solid to answer all the CF enigmas.

*Critique:* If one proceeds from habitual thinking drawn from conventional nuclear and plasma physics, the idea of multibody fusion appears wild to say the least. However, one has to take into account that the situation with CF is drastically different due to the presence in the solid of both periodically ordered positions for embedded d's and preferred directions for the motion of nuclei. This makes multibody collisions in principle more likely than in free space. However, no firm calculations have yet been presented which are testable. Quite independent of CF, interesting and unusual results have been obtained in channeling experiments of energetic particles in solids (Sorenson and Uggerhoj, 1987).

The multibody hypothesis carries along two other aspects which should be commented upon. One aspect is related to cluster-formation, strong electron screening by coherent dynamics, resonances, etc. At this stage, these corallary ideas are even more speculative than the basic multibody hypothesis. They are peripheral and neither make nor break the main idea. The second aspect, is that of nuclear transmutation where not only the H isotopes participate, but also the lattice nuclei. We note that there is as yet no theoretically well-founded model for such processes.

### 4.6 Exotic Chemistry

A number of models assume the existence of an exotic chemical system whose occurrence either precedes nuclear synthesis or makes it quite unnecessary. Such systems are assumed to be engendered by electromagnetic or nuclear interactions. The similarity of these postulated models is in their tight binding of electrons in atoms and/or molecules; with detailed analyses by Rice et al (1993).

One of the simplest (Mills and Farrell, 1990; and Mills and Kneizys, 1991), also deviates the least from the conventional view. These authors claim that in addition to the normal energy levels with a ground state of - 13.53 eV for the H atom, a more tightly bound sub-ground state of -27.21 eV is possible. This is obtained from a combination of a quasi-Laplace's equation and semi-classical physics in which the electron orbits are represented as spherical shells of uniformly distributed charge. For them the excess power, with no nuclear products, is simply the extra 13.68 eV/atom obtained as H isotopes go into the sub-ground state. About 60% of their 1991 paper elucidates their theory, with the remaider presenting their experimental results to justify it.

Maly and Va'vra (1993) increase the complexity by doing a calculation for the hydrogen atom based upon the relativistic Dirac equation and get an extremely tightly bound electron orbit. They get a binding energy ~ 500 keV, and a radius of ~ $5 \times 10^{-13}$ cm, a nuclear dimension. For them CF also involves no nuclear process. The excess energy is 500 keV/atom as these tightly bound atoms are formed. They suggest that this chemical ash of tightly bound H or D atoms may account for the missing mass (dark matter) of the universe. Such tightly bound H atoms would escape most planets and stars, and be stable unlike free n's.

The next set of models involve tight H isotope molecules in which excess energy may result chemically, and/or from nuclear fusion as the tightly bound atoms more easily penetrate their common Coulomb barrier. Cerofolini and Re (1990), assume the existence of a tightly bound D-D molecule of radius r <~ 0.25 A. They consider this to be one of many possible "binuclear atoms" in which "all the electrons rotate around the two bare nuclei." For a system of two D's, the arguments are based on the energy dependence for two extreme cases -- for short and for long distances.

At long distances, this energy is expressed as a sum of screened repulsion energy and the energy of ionization of two atoms (their zeroth-order energy corresponding to unbound electrons and d's):

$$E_{(D^+)e^-} + E_{(D^+)e^-} = \frac{e^2}{r}\exp(-\frac{r}{a}) - 2E_o. \tag{49}$$

At short distances, the system is perceived to be a binuclear helium-like atom, $(D^+D^+)2e^-$. For the energy of such a state, the following equations are proposed:

$$E_{dd2e}(r) = \frac{e^2}{r} - E_{elec}(r) , \text{ and} \tag{50}$$

$$E_{elec}(r) = [E_{elec}(0) - 2E_o]F(r) + 2E_o . \tag{51}$$

Here the function F(r) is an empirical one matching the limits: $\lim_{r \to 0} F(r) = 1$, and $\lim_{r \to \infty} F(r) = 0$. The quantity $E_{elec}(0)$ corresponds to two different energy states of a helium atom: $E_{el}(0) = 78.98$ eV for para-helium, and $E_{el}(0) = 59$ eV for ortho-helium. The function F(r) is selected by analogy with the theory for the $H_2^+$ ion.

While constructing the plot of the function E(r) from eq.(49) for $r > 0.5\ a_o$, and from eq.(50) for the region $r \leq 0.5\ a_o$, they note a decrease in the intermediate region between the left and right branches of the curve. This leads them to surmise the possible existence of a metastable state in this region. Figure 4 schematically shows such a hypothesized metastable state.

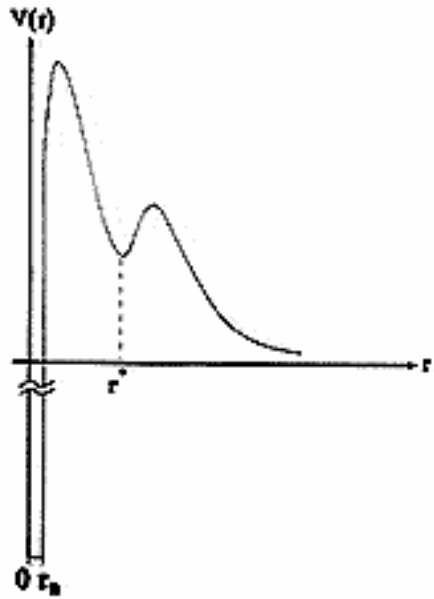

**Figure 4.** The hypothetical metastable state of two deuterons of Cerofolini and Re in Section 4.6 is depicted permitting them to tunnel through a narrowed part of the Coulomb barrier.

Cerofolini and Re think that in electrolytic and gas-loading experiments the mechanism of (DD)* production can be related to some phase transition types. For example in the chemical reaction $LiD + D_2O \oslash LiOD + D_2$, they point out that this process could lead to the formation of (DD)*. The electrolytic reaction $2D^+ + 2e \oslash D_2 + 31.7$ eV, that can proceed during the course of sorption and desorption at the cathode surface, is suggested as a source of the energy required for the formation of a metastable precursor.

Gryzinsky (1990) and Barut (1990) present analyses to substantiate the existence of the metastable $D_2^+$ state based on three body calculations for two d's and one electron. Gryzinsky treats the problem mainly classically, but neglects radiation effects for his oscillating electron as allowed by quantum mechanics. Barut's analysis is based on the Bohr-Sommerfeld quantization principle, and obtains a binding energy of 50 keV. Both authors, independently, conclude that a "superbound" $(D_2^+)^*$ molecular ion can exist in which an electron that is exactly half-way between the d's provides an attractive force and screens the d Coulomb repulsion. Vigier (1992) presents an analysis almost identical to that of Barut.

Mayer and Reitz (1991) claim that resonances of ep, ed, and et are created which if the fleeting unions survive long enough, allow a high probability of barrier penetration and subsequent nuclear reaction.

*Critique:* Because the Mills et al results differ only a little from the conventional, they should be easy to test experimentally. They only need to demonstrate the spectral line of the first excited state above their sub-ground state. This has not been done. If their tight H abounds in the universe, one may ask why this spectral line has not been seen long ago. For all the models, one may raise the general question: If tight orbits are possible for H isotopes, why not for all atoms; and why haven't such spectral lines been seen ?

Because the Maly et al result differs so much from conventional, it is much harder to test experimentally in the same way as for Mills et al, because a much higher excitation energy of $\sim$ 500 keV rather than $\sim$ 10 eV is needed. Also these extremely tight H atoms would mimic neutrons and be difficult to trap and contain for testing. In fact with orbits $\sim 5 \times 10^{-13}$ cm = 5 fm, these are tighter than 250 fm muonic orbits and should produce fusion at a much higher rate than muon catalyzed fusion. The missing nuclear ash problem is thus not eliminated. However there are also some serious errors in their analysis. At the nuclear surface, $r = r_n$  0, and hence both regular and irregular solutions are simultaneously allowed for $r$  $r_n$. Therefore, a general solution is a linear combination of them for $r$  $r_n$. When the boundary conditions are imposed at $r = r_n$, it can be shown that the irregular component becomes nearly negligible compared to the regular component (Rice et al, 1994). Thus the results of Maly et al are incorrect, since they assumed erroneously that the general solution can be given by just the irregular solution, independent of the regular solution.

Actually there seems to be no basis for assuming the existence of a binuclear atom $(D^+D^+)2e^-$, or a a superbound state of the $(D_2^+)^*$ ion. Some of the Cerofolini analysis is qualitative, and the V(r) curves are only approximately represented. The most critical length region is beyond and at best barely at the boundary of applicability of the equations. It would be more appropriate to replace the curves with bands to account for the uncertainty in the description.. The band width increases as the boundary of applicability of the empirical equations is approached. The true curve should lie inside the band width. An exact solution for the entire region under consideration will likely yield a smooth curve containing no local minimum. Thus the metastable state may not be present in the more rigorous analysis (Kolos,1986). Therefore the "superbound" solution is at best unstable.

Resonant ep, ed, and et states seem not to be supported by existing data. The claims of Mayer et al (1991) are based on the resonance models of Spence et al (1990,1991) and Benesh et al (1990) who used single photon exchange in the Coulomb gauge for the ep system. Recently, McNeil (1991) reformulated the ep problem in a qualitative yet gauge-invariant way, and found no evidence for a resonance in the ep system in the low energy range of interest. Therefore neither experiment nor theory seem to support such resonances.

For Barut, Gryzinsky, and Vigier, the analysis is predicated on very unlikely precise symmetry. The electron must be exactly between the two d's and the system is unstable. The tightness of the orbit appears to violate the uncertainty principle. Although a non-relativistic analysis is warranted for slow, large mass H isotopes around the electron, a non-stationary electron will require a relativistic treatment because it will attain a velocity close to the velocity of light due to its small mass. Perhaps a full relativistic calculation including spin-spin and spin-orbit coupling may save this model, but this has not been presented as yet. Their electron orbits ~ 20 fm, are 13 times smaller than 250 fm muonic orbits and would produce fusion at a much higher rate than muon catalyzed fusion.

## 5. CONCLUSION

We conclude that in spite of considerable efforts, no theoretical formulation of CF has succeeded in quantitatively or even qualitatively describing the reported experimental results. Those models claiming to have solved this enigma appear far from having accomplished this goal. Perhaps part of the problem is that not all of the experiments are equally valid, and we do not always know which is which. We think that as the experiments become more reliable with better equipment etc., it will be possible to establish the phenomena, narrow down the contending theories, and zero in on a proper theoretical framework; or to dismiss CF. There is still a great deal of uncertainty regarding the properties and nature of CF.

Of course, the hallmark of good theory is consistency with experiment. However, at present because of the great uncertainty in the experimental results, we have been limited largely in investigating the consistency of the theories with the fundamental laws of nature and their internal self-consistency. A number of the theories do not even meet these basic criteria.

Some of the models are based on such exotic assumptions that they are almost untestable, even though they may be self-consistent and not violate the known laws of physics. It is imperative that a theory be testable, if it is to be considered a physical theory.

The simplest and most natural subset of the theories are the acceleration models. They do explain a number of features of the anomalous effects in the deuterated systems. However these models seem incapable of explaining the excess energy release which appears to be uncorrelated with the emission of nuclear products; and incapable of explaining why the branching ratio t/n >>1. If these features continue to be confirmed by further experiments, we shall have to reject the acceleration mechanism also.

It is an understatement to say that the theoretical situation is turbid. We conclude that the mechanism for anomalous effects in deuterated metals is still unknown. At present there is no single consistent theory that predicts or even explains CF and its specific features from first principles.

*Since The Third Int'l. Conf. on Cold Fusion held in Nagoya, Japan on Oct.21-25, 1992 is frequently cited, it is designated as *ICCF* **3**.